\documentclass[aps,preprint,11pt,amsmath,amssymb,superscriptaddress,floatfix]{revtex4-1}
\usepackage[colorlinks,citecolor=blue,linkcolor=blue]{hyperref}
\usepackage{graphicx}
\usepackage{dcolumn}
\usepackage{bm} 
\usepackage[utf8]{inputenc}
\usepackage{txfonts}

\newcommand{\JGU}{Institut f{\"u}r Physik, Johannes Gutenberg Universit{\"a}t Mainz, D-55099 Mainz, Germany}
\newcommand{\UPP}{Department for Physics and Astronomy, Uppsala University, SE-75105 Uppsala, Sweden}
\newcommand{\CAU}{Institute of Theoretical Physics and Astrophysics, Christian-Albrechts-Universit{\"a}t zu Kiel, D-24098 Kiel, Germany}
\newcommand{\CNN}{Centre de Nanosciences et de Nanotechnologies, CNRS, Univ. Paris-Sud, Universit{\'e} Paris-Saclay, 91405 Orsay, France}
\newcommand{\PRA}{Institute of Physics, Academy of Sciences of the Czech Republic, Cukrovarnick\'{a} 10, 162 53 Praha 6 Czech Republic}

\begin{document}

\title{Trochoidal motion and pair generation in skyrmion and antiskyrmion dynamics under spin-orbit torques}
\author{Ulrike Ritzmann}
\affiliation{\JGU}\affiliation{\UPP}
\author{Stephan von Malottki}
\affiliation{\CAU}
\author{Joo-Von Kim}
\affiliation{\CNN}
\author{Stefan Heinze}
\affiliation{\CAU}
\author{Jairo Sinova}
\affiliation{\JGU}\affiliation{\PRA}
\author{Bertrand Dupé}
\affiliation{\JGU}

\date{\today}

\begin{abstract}
Skyrmions and antiskyrmions in magnetic ultrathin films are characterised by a topological charge describing how the spins wind around their core. This topology governs their response to forces in the rigid core limit. However, when internal core excitations are relevant, the dynamics become far richer.  We show that current-induced spin-orbit torques can lead to phenomena such as trochoidal motion and skyrmion-antiskyrmion pair generation that only occurs for either the skyrmion or antiskyrmion, depending on the symmetry of the underlying Dzyaloshinskii-Moriya interaction. Such dynamics are induced by core deformations, leading to a time-dependent helicity that governs the motion of the skyrmion and antiskyrmion core. We compute the dynamical phase diagram through a combination of atomistic spin simulations, reduced-variable modelling, and machine learning algorithms. It predicts how spin-orbit torques can control the type of motion and the possibility to generate skyrmion lattices by antiskyrmion seeding. 
\end{abstract}

\maketitle

Two-dimensional spin structures such as vortices and skyrmions~\cite{Bogdanov1989a, Bogdanov1999} possess a nontrivial topology that affords them a degree of stability~\cite{Hagemeister2015, Rohart2016, Stosic2017}. These structures are characterised by a topological winding number or `charge',
\begin{equation}
q = - \frac{1}{4 \pi} \int d^2r \; \mathbf{m} \cdot \left( \frac{\partial \mathbf{m}}{\partial x} \times \frac{\partial \mathbf{m}}{\partial y} \right),
\end{equation}
where $\mathbf{m} = \mathbf{m}(\mathbf{r},t)$ is a unit vector representing the orientation of the magnetic moments in time and space. Skyrmions ($q = 1$) and antiskyrmions $(q = -1)$, for example, possess opposite charges and can appear in pairs through the continuous deformation of the uniform state ($q=0$)~\cite{Koshibae2016, Everschor-Sitte2016, Stier2017}. The description of the dynamics of skyrmions and antiskyrmions can be approximated by assuming a rigid core, which leads to a reduced set of variables describing their motion. This dynamics is captured by the Thiele equation~\cite{Thiele1973},
\begin{equation}
\mathbf{G} \times \frac{\partial \mathbf{X}}{\partial t} + \alpha D_0 \frac{\partial \mathbf{X}}{\partial t}  = \mathbf{F}.
\label{eq:Thiele}
\end{equation}
which describes the damped gyrotropic motion of the (anti-)skyrmion core position, $\mathbf{X}$, in response to a force $\mathbf{F}$. Here $\mathbf{G} = -q \, G_0 \, \hat{\mathbf{z}}$ is the gyrovector, $\alpha$ is a damping constant, and $D_0$ is a structure factor related to the damping (see Methods). While the dynamics in Eq.~\ref{eq:Thiele} is non-Newtonian, the gyrotropic response depends on $q$, i.e., its topology, and dictates the direction in which the core moves. This conceptual framework has been useful to understand vortex dynamics~\cite{Guslienko2002, Choe2004}, spin-torque vortex oscillators~\cite{Ivanov2007, Mistral2008}, and the current-driven motion of skyrmions~\cite{Sampaio2013, Nagaosa2013, Lin2013b, Koshibae2016, Lin2016a, Everschor-Sitte2016, Leonov2016c}.

In most studies to date, however, the robustness of the symmetry between opposite topological charges, as expressed in Eq.~(\ref{eq:Thiele}), has not been examined in detail. In particular, the roles of core deformation beyond inertial effects~\cite{Buttner2015}, the internal degrees of freedom, and the underlying symmetry of the magnetic interactions that stabilise the skyrmions remain an open question. This issue is of particular importance since nanometre-scale skyrmions are desirable for possible device applications~\cite{Fert2013, Fert2017} and antiskyrmions have been observed in Heusler compounds~\cite{Nayak2017} and predicted to occur at transition metal interfaces~\cite{Hoffmann2017}. We show here that the symmetries of the magnetic interactions, combined with spin-orbit torques (SOTs), play an important role in determining how the (anti)skyrmion core moves. In particular, the choice of the Dzyaloshinskii-Moriya interaction (DMI) can lead to \emph{qualitatively different} motion for opposite $q$ charges. Namely, deviations from rectilinear motion and skyrmion-antiskyrmion pair generation can occur above certain SOT thresholds for the skyrmion \emph{or} the antiskyrmion depending on the choice of DMI.

\begin{figure}
\centering\includegraphics[width=8cm]{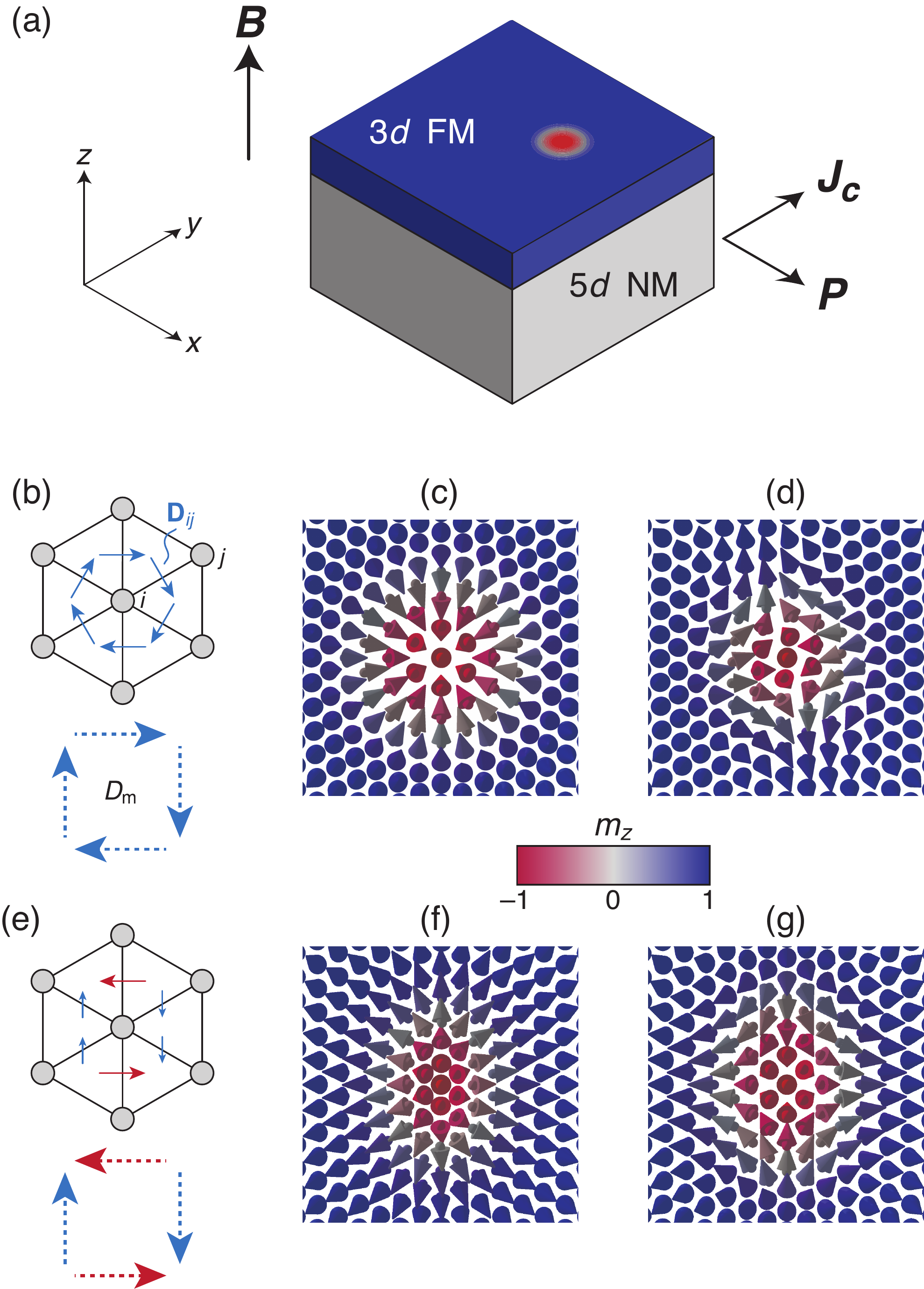}
\caption{Film geometry, symmetry of the DMI, and skyrmion profiles. (a) Bilayer system studied comprising an ultrathin 3\emph{d} transition metal ferromagnet (FM) on a 5\emph{d} normal metal (NM) substrate, with the configuration of the applied magnetic field ($\mathbf{B}$), charge current ($\mathbf{J_c}$), and effective spin polarization vector ($\mathbf{P}$). (b) Hexagonal lattice structure and orientation of the DMI $\mathbf{D}_{ij}$ used in the atomistic spin dynamics simulations. The dashed arrows represent the effective Dzyaloshinskii-Moriya vectors $\mathbf{D}_m$ in the continuum (micromagnetic) limit. (c) Equilibrium skyrmion ($q=1$) configuration with the DMI in (b). (d) Equilibrum antiskyrmion ($q=-1$) configuration with the DMI in (b). (e) Hexagonal lattice structure and orientation of the modified DMI vectors used to favour the antiskyrmion state. The dashed arrows represent the effective Dzyaloshinskii-Moriya vectors $\mathbf{D}_m$ in the continuum limit.  (f) Equilibrium skyrmion ($q=1$) configuration with the DMI in (e). (g) Equilibrium antiskyrmion ($q=-1$) configuration with the DMI in (e).}
\label{fig:geometry}
\end{figure}

To explore this issue in greater depth, we studied theoretically the spin dynamics of skyrmions and antiskyrmions in an ultrathin 3\emph{d} transition metal ferromagnet on a 5\emph{d} normal metal substrate as shown in Figure~\ref{fig:geometry}(a). 
A prominent example of this material combination is PdFe/Ir(111), where a large DMI is induced in the Fe monolayer through interfacial coupling to the strong spin-orbit interaction in the Ir substrate~\cite{Fert1980, Dupe2014}. This allows individual skyrmions to exist as metastable states, which has been brought to light in recent experiments~\cite{Romming2013}. Moreover, it has been shown that a variety of antiskyrmion states ($q=-1,-2$) are also metastable when frustrated exchange interactions are taken into account in such ultrathin films~\cite{Dupe2016a, Boettcher2017, VonMalottki2017, Leonov2015} and more generally in bulk chiral magnets~\cite{Zhang2017, Hu2017}, which lead to an attractive interaction between skyrmions~\cite{Rozsa2016a}. We employed density-functional theory calculations to obtain estimates of the exchange, anisotropy, and DMI energies for PdFe/Ir(111), which were then used to parametrise an atomistic spin model for studying the dynamics (see Methods). Minimising this energy allows us to determine the equilibrium spin configuration of the static skyrmion and antiskyrmion profiles, as shown in Figure~\ref{fig:geometry}(c) and \ref{fig:geometry}(d), respectively. Note that the exchange and DMI possess a six-fold symmetry that is consistent with the Ir(111) surface [Fig.~\ref{fig:geometry}(b)].

\begin{figure}
\centering\includegraphics[width=12cm]{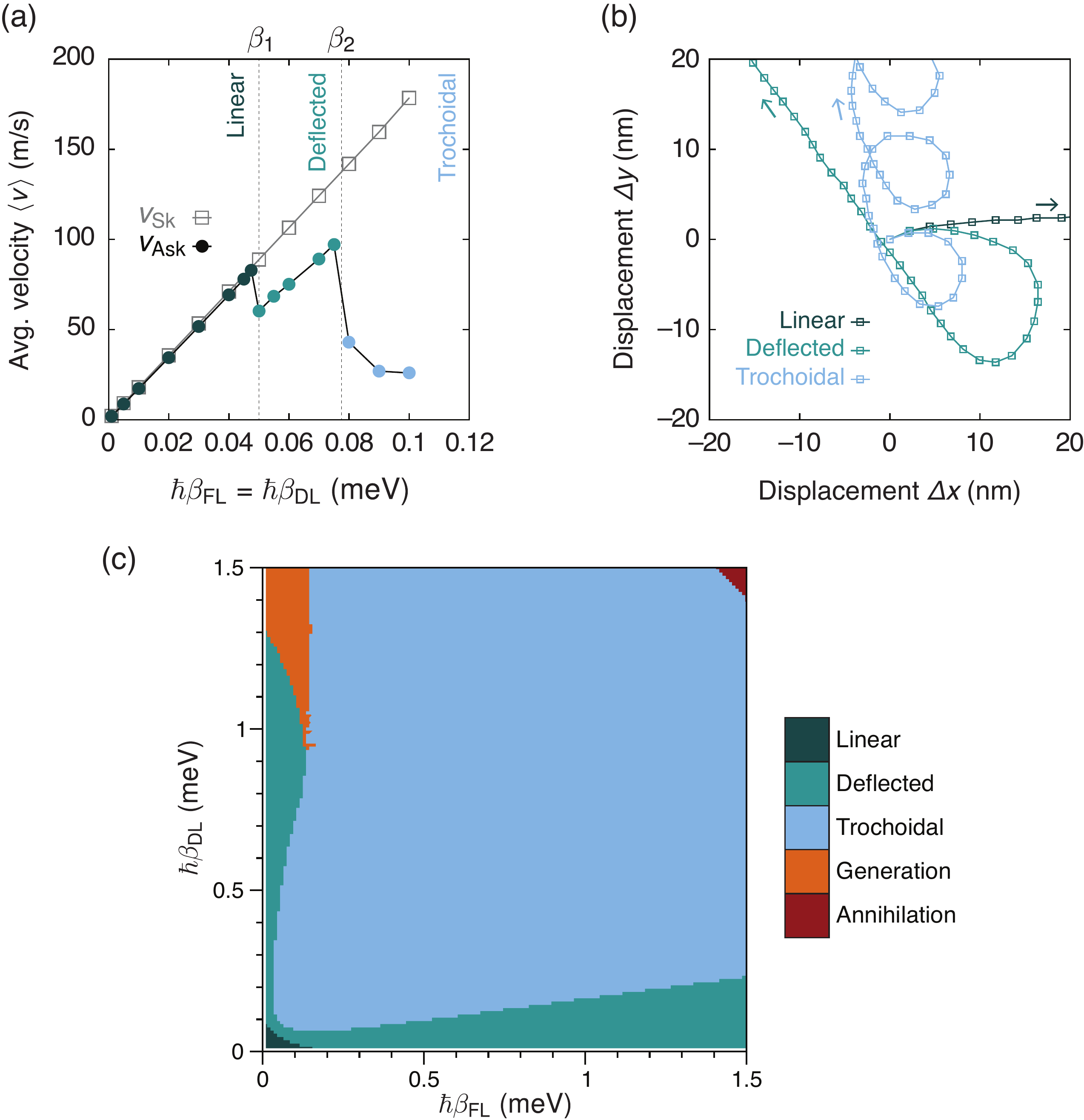}
\caption{Motion of antiskyrmions under current induced spin transfer torques. (a) Variation of the average velocity $\langle v \rangle$ of skyrmions and antiskyrmions as a function of SOT, where $\beta_\mathrm{FL}=\beta_\mathrm{DL}$. Three propagation regimes are identified for antiskyrmions:  rectilinear motion at low currents, deflected motion at intermediate currents, and trochoidal motion at high currents. (b) Example of antiskyrmion trajectories for linear ($\hbar\beta_\mathrm{FL}=\hbar\beta_\mathrm{DL}=0.04$meV), deflected ($\hbar\beta_\mathrm{FL}=\hbar\beta_\mathrm{DL}=0.06$meV) and trochoidal motion ($\hbar\beta_\mathrm{FL}=\hbar\beta_\mathrm{DL}=0.09$meV). The arrows indicate the propagation direction. (c) $\beta_\mathrm{FL}-\beta_\mathrm{DL}$ phase diagram for the antiskyrmion dynamics.}
\label{fig:askdyn}
\end{figure}

The spin dynamics are computed by time integrating the Landau-Lifshitz-Gilbert equation with additional SOT terms due to the applied current,
\begin{equation}
	\frac{d \mathbf{m} }{dt}= -\frac{1}{\hbar} \mathbf{m} \times \mathbf{B_{\mathrm{eff}}} + \alpha \mathbf{m} \times \frac{d \mathbf{m} }{dt} + \beta_\mathrm{FL} \mathbf{m} \times \mathbf{P}  + \beta_\mathrm{DL} \mathbf{m} \times \left( \mathbf{m} \times \mathbf{P} \right) ,
	\label{eq:LLG}
\end{equation}
where $\hbar$ is the Planck constant, $\mathbf{B}_{\mathrm{eff}}=-\delta H/\delta \mathbf{m}$ is the effective field, $\alpha$ is the Gilbert damping constant, $\beta_\mathrm{FL}$ is the strength of the field-like torque, and $\beta_\mathrm{DL}$ is the strength of the damping-like torque. $\mathbf{P} = \hat{\mathbf{x}}$ is the orientation of the effective spin polarisation, which models an applied electric current along the $y$ direction in the film plane [Fig.~\ref{fig:geometry}(a)]. While in-plane currents should in principle flow through both the ferromagnet and normal metal substrate, we assume that the majority of the current flows only through the substrate since the layer resistivity is significantly larger for the ultrathin ferromagnet (one or two monolayers thick), given the importance of interfacial scattering~\cite{Sondheimer2001} and its relative thickness in comparison to the substrate. We can therefore neglect spin transfer torques generated within the 3\emph{d} ferromagnet and assume only field-like and damping-like contributions from the spin-orbit coupling in the 5\emph{d} substrate. An example of the ensuing current-driven motion of a skyrmion is shown in Fig.~\ref{fig:askdyn}a, where the average velocity is plotted as a function of the SOT for the case of $\beta_\mathrm{FL} = \beta_\mathrm{DL}$. This behaviour is consistent with the Thiele equation, which predicts a linear variation of the skyrmion velocity as a function of the SOT~\cite{Sampaio2013}.

In contrast to skyrmions, the average velocity of antiskyrmions does not increase monotonically with the SOT [Fig.~\ref{fig:askdyn}(a)]. 
A linear regime is found at low currents up to a first threshold, $\beta_1$, where a discontinuity in the  velocity curve can be seen. Above this threshold, the velocity continues to increase linearly as a function of $\beta$ but with a different slope. A second threshold $\beta_2$ is found as the strength of the SOT is increased, where the velocity decreases with the applied current. The calculated trajectories for the antiskyrmion core are presented in Fig.~\ref{fig:askdyn}(b). For linear motion, we observe that the spin configuration of the core is slightly deformed but remains close to its equilibrium static configuration. Above the first threshold $\beta_1$, the trajectory is linear at long times but exhibits a large transient phase in which the motion is curved. The rotation ceases when a new steady state regime is reached, which then allows for linear motion to proceed indefinitely (albeit with a different Hall angle with respect to the linear case $\beta < \beta_1$). More interestingly, the core undergoes trochoidal motion for $\beta > \beta_2$  which comprises an average displacement along a line that is accompanied by oscillations resulting in loops along the trajectory. The onset of these oscillations results in the sharp decrease in the average velocity shown in Fig.~\ref{fig:askdyn}(a). The phase diagram of the different behaviour is shown in Fig.~\ref{fig:askdyn}(c) for different values and ratios of $\beta_\mathrm{FL}$ and $\beta_\mathrm{DL}$. We used algorithms based on machine learning to classify the three types of trajectories (linear, deflected, and trochoidal), which exhibit a wide range of velocities and propagation directions (see Methods). We note that the trochoidal motion occurs over a wide range of SOT parameters.

\begin{figure}
\centering\includegraphics[width=10cm]{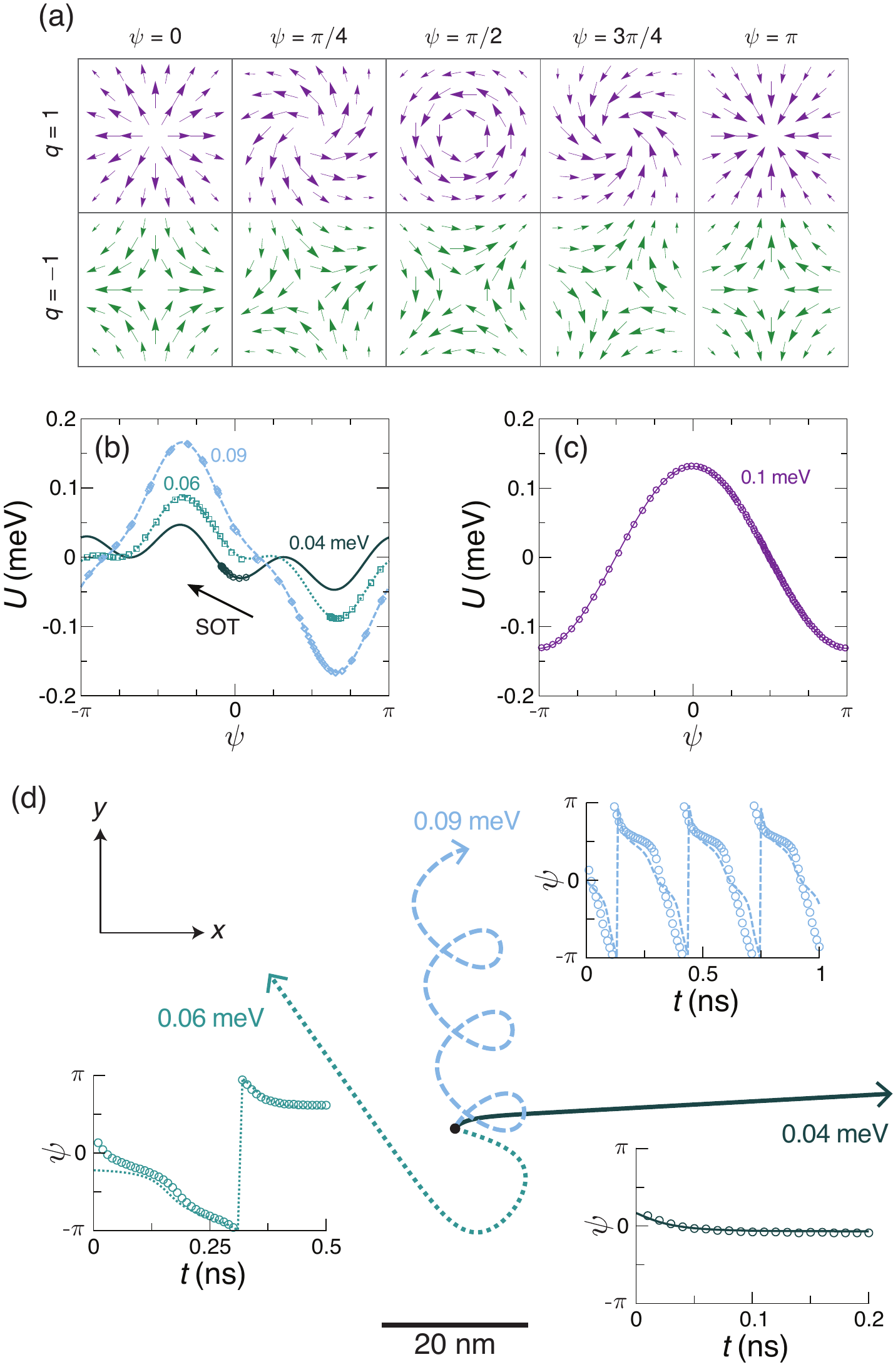}
\caption{Helicity dynamics in the extended Thiele model. (a) Spin configurations at the skyrmion ($q=1$) and antiskyrmion ($q=-1$) core for different values of the helicity parameter, $\psi$. (b) The antiskyrmion energy $U(\psi)$ for different values of $\beta_\textrm{FL} = \beta_\textrm{DL}$, where the figures denote the SOT strength in meV. Three regimes are shown: linear (0.04 meV), deflected (0.06 meV), and trochoidal (0.09 meV). Open symbols represent data extracted from spin dynamics simulations, while lines represent fits to the function $U(\psi) = u_1 \cos(\psi-\psi_0)+u_2 \cos(3\psi)$. The arrow indicates schematically the SOT force.  (c) $U(\psi)$ for a skyrmion with the DMI constant $D_{ij}$ reduced by a factor of $10^3$, for which the trochoidal regime is attained for $\hbar \beta_\mathrm{FL} =  \hbar \beta_\mathrm{DL} = 0.01$ meV. (d) Trajectories for $\beta_\mathrm{FL} = \beta_\mathrm{DL}$ using Eqs.~(\ref{eq:Thiele}) and (\ref{eq:Thielepsi}) with the fits for $U(\psi)$ in (b). The insets show $\psi(t)$ extracted from simulations (circles) and computed using Eq.~(\ref{eq:Thielepsi}) with the fitted $U(\psi)$ in (b) (lines).}
\label{fig:model}
\end{figure}

The deflected and trochoidal motion are driven by deformations to the antiskyrmion core. This deformation is characterised by the emergence of a dynamical variable $\psi(t)$ that describes the helicity of the skyrmion and antiskyrmion [Fig.~\ref{fig:model}(a)]. 
For skyrmions, $\psi$ describes the continuous transition between Bloch and N\'eel states of opposite chirality, while for antiskyrmions it describes the rotation of the Bloch or N\'eel axes. In our system, the deformation is driven by the SOT, which results in a tilt in the magnetization in the film plane, characterised by an amplitude $\eta$ and the azimuthal angle $\phi_t$, that depends on the relative strength between the field-like and damping-like terms. This tilt is uniform for the background spins, while it varies within the antiskyrmion core depending on the orientation of $\psi$. By assuming a suitable \emph{ansatz} for the deformation profile (see Methods and Supplementary Information), we can derive an equation of motion for $\psi(t)$ using a Lagrangian approach,
\begin{equation}
D_{\psi} \frac{\partial \psi}{\partial t} = \sigma_{\psi} \,  \hbar \beta_\textrm{DL} \, \eta \cos\left(\phi_p - \phi_t\right) - \frac{\partial U}{\partial \psi},
\label{eq:Thielepsi}
\end{equation}
where $D_{\psi}$ is a damping structure factor, $\sigma_{\psi}$ is an SOT efficiency factor, $\phi_p$ is the azimuthal angle of the spin polarisation vector $\mathbf{P}$ ($\phi_p = 0$ in the simulations), and $U$ is the internal magnetic energy. Because the effective SOT force acting on $\mathbf{X}(t)$ in Eq.~(\ref{eq:Thiele}) can be written as
\begin{equation}
\mathbf{F} = \sigma_0 \, \hbar \beta_\textrm{DL} \left( \sin\left(\psi-\phi_p \right), \cos\left(\psi-\phi_p \right) \right),
\end{equation}
the dynamics of $\psi$ determines the time dependence of the force and therefore the overall trajectory of the antiskyrmion as shown in Fig~\ref{fig:askdyn}(b), and results from the interplay between the SOT term $\sigma_\psi$ and the restoring force governed by $\partial_\psi U$. We verified this interpretation by computing the spatially-resolved forces from the atomistic spin simulations (see Supplementary Information). In Fig.~\ref{fig:model}(b) and \ref{fig:model}(c), we present $U(\psi)$ extracted from the spin dynamics simulations for different SOT strengths. We find that the potential can be described accurately by the function $U(\psi) = u_1 \cos(\psi-\psi_0) + u_3 \cos(3\psi)$, where $u_1 \propto \eta^2$ and $\psi_0 \propto 2\phi_t$ for antiskyrmions, which is consistent with predictions from the model (see Methods). The $u_3$ term represents a lattice effect that accounts for the underlying hexagonal lattice structure~\cite{Rozsa2017} and is found to be largely independent of the SOT. The position of the energy minimum is largely independent of the SOT for $\beta_\textrm{FL} = \beta_\textrm{DL}$ because the tilt $\phi_t$ remains almost constant for this torque ratio [Fig.~\ref{fig:model}(b)]. However, cases where $\beta_\textrm{FL} \neq \beta_\textrm{DL}$ lead to different values of $\phi_t$, which results in a shift in the minimum (see Supplementary Information).

From $U(\psi)$, we can understand the salient features of Eq.~(\ref{eq:Thielepsi}) as follows. For low amplitudes of the SOT, the restoring force due to the lattice term $u_3$ dominates and the steady state value of $\psi$ remains close to its equilibrium value $\psi \simeq 0$, resulting in the simple linear motion expected from Eq.~(\ref{eq:Thiele}) alone. As the strength of the SOT is increased, the deformation-induced contribution $u_1$, which is also governed by the DMI, increases and leads to a change in stability, where a new steady state value $\psi = \psi_0$ is reached. This results in the deflected motion, which is characterised by large transients in $\psi(t)$ leading to the stationary value $\psi_0$ at long times. As the SOT is further increased, the trochoidal regime is attained when the SOT contribution exceeds the maximum value of the restoring force $|\partial_\psi U|$, which results in a periodic solution in $\psi(t)$. In this light, the transition toward the trochoidal regime is analogous to Walker breakdown in domain wall motion~\cite{Slonczewski1973}, where the magnetization angle at the domain wall centre plays the role of $\psi$ here. By using the fits in Fig.~\ref{fig:model}(b), we computed the dynamics of $\psi(t)$ using Eq.~(\ref{eq:Thielepsi}) to determine the antiskyrmion trajectories in the three regimes [Fig.~\ref{fig:model}(d)]. We note that the predicted dynamics of $\psi(t)$ and $\mathbf{X}(t)$ accurately reproduce the behaviour obtained from the atomistic spin dynamics simulations described by Eq.~(\ref{eq:LLG}). These results also illustrate why such transitions are not seen for the skyrmion under similar conditions; Fig.~\ref{fig:model}(c) shows that similar variations in $U(\psi)$ can only be obtained for a skyrmion if the DMI constant is reduced by a factor of $10^3$, which indicates that the equilibrium skyrmion state is robust and remains largely unperturbed under SOT in this geometry.

\begin{figure}
\centering\includegraphics[width=12cm]{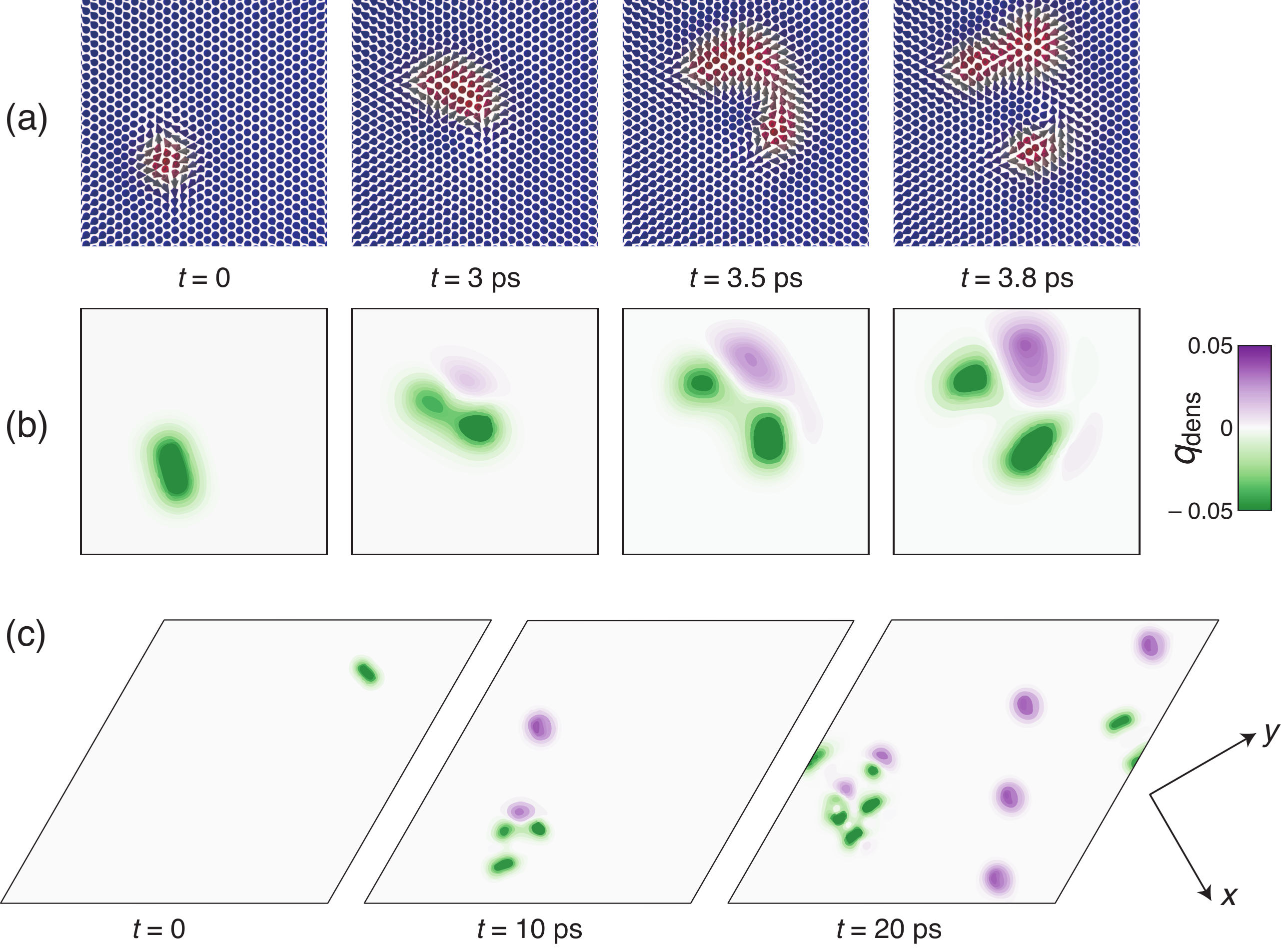}
\caption{Skyrmion-antiskyrmion pair generation from trochoidal antiskyrmion dynamics for $\hbar\beta_\mathrm{FL}=0.01$ meV and $\hbar\beta_\mathrm{DL}=1.35$ meV. (a) Snapshots of the trochoidal motion of a single antiskyrmion, where large deformations in the spin structure of the core leads to the nucleation of a skyrmion-antiskyrmion pair. (b) Topological charge density $q_\mathrm{dens}$ corresponding to the spin states in (a). (c) Snapshots in time of skyrmion-antiskyrmion pair generation, where pairs are nucleated and annihilated periodically. Antiskyrmions that survive become subsequent sources of pair generation.}
\label{fig:pairgen}
\end{figure}

Two other regimes beyond the single-particle description are also identified in Fig.~\ref{fig:askdyn}(c). First, under large field-like and damping-like torques, the propagating antiskyrmion is no longer stable and becomes annihilated. Second, and more interestingly, a transition toward another dynamical regime is found under small $\beta_\mathrm{FL}$ and large $\beta_\mathrm{DL}$, where deflected or trochoidal motion leads to a periodic generation of skyrmion-antiskyrmion pairs. An example of this process is given in Figure~\ref{fig:pairgen}. 
This regime is strongly nonlinear and represents a complete breakdown of the single-particle picture described by Eqs.~\ref{eq:Thiele} and \ref{eq:Thielepsi} (see Supplementary Information). The pair is generated as follows: As the antiskyrmion undergoes its trochoidal trajectory, it is accompanied by a large deformation which represents an elongation of the core [i.e., at $t=3$ ps in Fig.~\ref{fig:pairgen}(a)], similar to the dynamics seen for gyrating magnetic vortices close to the core reversal transition~\cite{VanWaeyenberge2006, Yamada2007, Gaididei2010}. This elongation, which represents a skyrmion-antiskyrmion pair with a net charge of $q=0$, then separates from the core itself ($t=3.5$ to $3.8$ ps). The corresponding topological charge density $q_\textrm{dens}$ for these processes is shown in Fig.~\ref{fig:pairgen}(b). Once nucleated, the pair itself separates since the SOTs lead to different motion for the skyrmion and antiskyrmion constituents. The skyrmion propagates away from the nucleation site by undergoing rectilinear motion, while the nucleated antiskyrmion executes trochoidal motion and becomes itself a new source of pair generation. This remarkable phenomenon leads to the generation of a gas of skyrmions and antiskyrmions [Fig.~\ref{fig:pairgen}(c)];  the relative population of the two species varies in time as collisions between skyrmions and antiskyrmions lead to annihilation, while pair generation continues for antiskyrmions that survive. This process suggests that it is possible to generate an indefinite number of skyrmions and antiskyrmions from a \emph{single} antiskyrmion `seed'. Combined with the attractive interaction between cores made possible by the frustrated exchange, this dynamics can eventually lead to a skyrmion `crystallite' that condenses from the disordered gas phase (see Supplementary Information). This behaviour is very different to skyrmion generation reported previously, where single pairs are nucleated from static defects through the coupling between local magnetization gradients and spin transfer torques, in systems where skyrmions or antiskyrmions are unstable (depending on the choice of DMI)~\cite{Everschor-Sitte2016, Stier2017}.

Recall that we only observed deflected and trochoidal motion for antiskyrmions because the energy barriers $U(\psi)$ in the helicity $\psi$ are orders of magnitude larger for skyrmions than for antiskyrmions for the same DMI constant (Fig.~\ref{fig:model}). As such, the asymmetry between opposite topological charges is related to the form of the underlying DMI, rather than the sign of the charge itself. To test this hypothesis we conducted simulations in which an anisotropic form of the DMI is used instead, whereby the original six-fold symmetry is retained for the exchange interactions while a two-fold symmetry is used for the DMI, the DMI strength along the two axes being different as shown in Fig.~\ref{fig:geometry}(e). This mimics the symmetry of the DMI induced at a (110) interface~\cite{Hoffmann2017}. Since the amplitudes of the magnetic interactions are unchanged but only the symmetry, the stability of the magnetic textures is only qualitatively affected. Most importantly, antiskyrmions are favoured energetically over skyrmions for this anisotropic form of the DMI.

\begin{figure}
\centering\includegraphics[width=12cm]{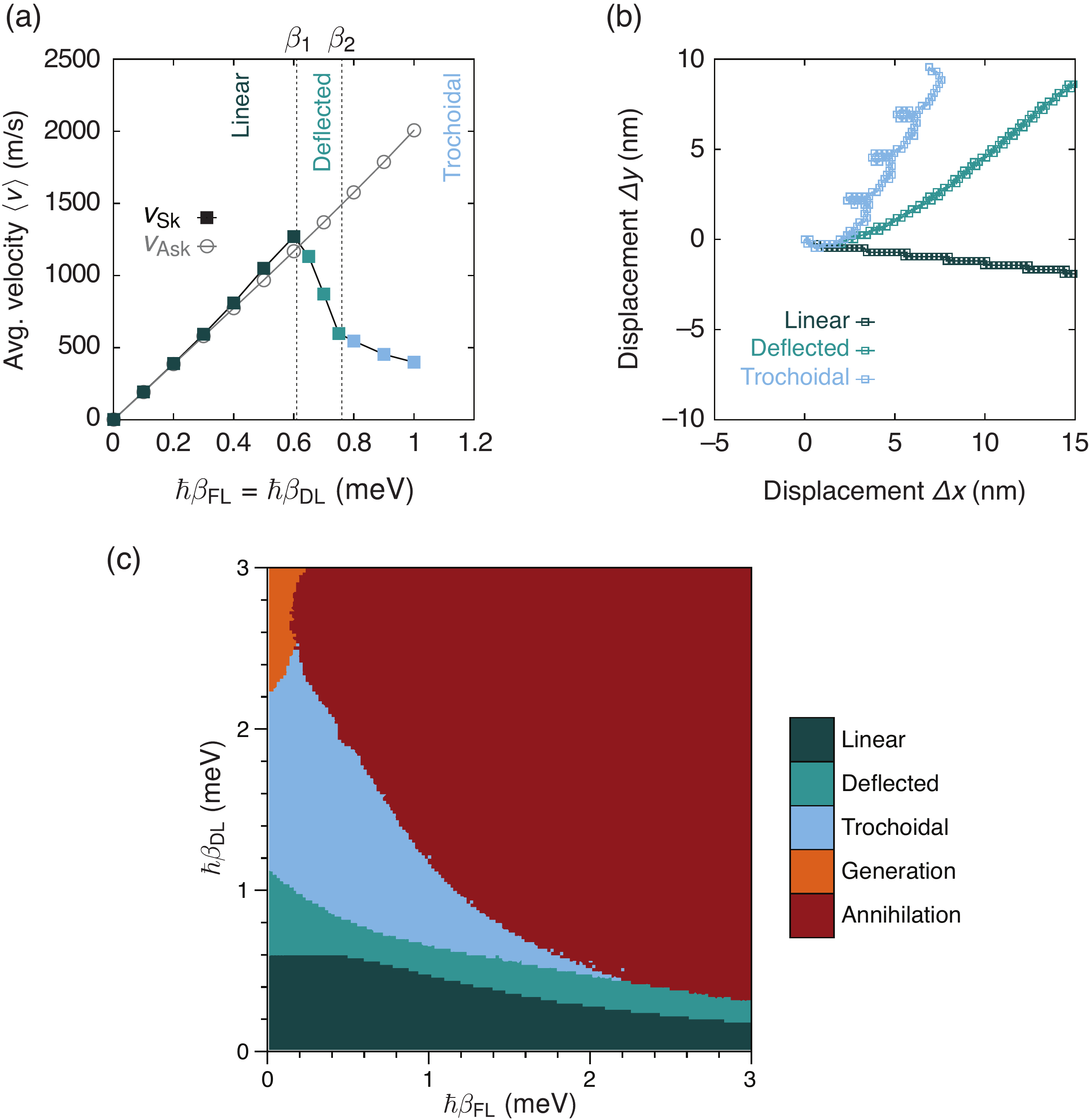}
\caption{Skyrmion and antiskyrmion dynamics with an anisotropic DMI. (a) Variation of the velocity of skyrmions and antiskyrmions as a function of the torques, where $\beta_\mathrm{FL}=\beta_\mathrm{DL}$, for the DMI shown in Fig.~\ref{fig:geometry}(e). (b) Example of skyrmion trajectories for linear ($\hbar\beta_\mathrm{FL}=\hbar\beta_\mathrm{DL}=0.5$meV), deflected ($\hbar\beta_\mathrm{FL}=\hbar\beta_\mathrm{DL}=0.7$meV) and trochoidal motion ($\hbar\beta_\mathrm{FL}=\hbar\beta_\mathrm{DL}=0.8$meV). (c) $\beta_\mathrm{FL}-\beta_\mathrm{DL}$ phase diagram for the skyrmion motion.}
\label{fig:anisDMI}
\end{figure}

Figure~\ref{fig:anisDMI} summarises the current-driven dynamics of skyrmions and antiskyrmions with the anisotropic DMI in Fig.~\ref{fig:geometry}(e).
In Fig.~\ref{fig:anisDMI}(a), the current-dependence of the velocity is shown for the skyrmion and antiskyrmion, whose static profiles are shown in Fig.~\ref{fig:geometry}(f) and \ref{fig:geometry}(g), respectively. In contrast to the behaviour shown in Fig.~\ref{fig:askdyn}, the antiskyrmion undergoes only rectilinear motion while the skyrmion exhibits deflected and trochoidal motion as the strength of the SOTs is increased. We note that the associated thresholds, $\beta_1$ and $\beta_2$, are also much higher, where velocities beyond 1 km/s can be reached in the linear regime for both skyrmions and antiskyrmions. In Fig.~\ref{fig:anisDMI}(b), examples of trajectories for the linear, deflected, and trochoidal motion for skyrmions are shown. We note that the overall skyrmion Hall angles are different to the antiskyrmion case shown in Fig.~\ref{fig:askdyn}(b), which originates from different stationary values of $\psi$ for the skyrmion. This is a consequence of $U(\psi)$ for the skyrmion with the anisotropic DMI, which possesses a different $\psi$-dependence than the case shown in Fig.~\ref{fig:model}(b). This difference is also reflected in the $(\beta_\mathrm{FL},\beta_\mathrm{DL})$ phase diagram in Fig.~\ref{fig:anisDMI}(c); while the same phases are identified, the overall shape of the phase boundaries differs and certain transitions are absent, such as the transition between deflected motion and pair generation. Nevertheless, the order in which the phases appear with increasing SOT is similar.

\begin{figure}
\centering\includegraphics[width=15cm]{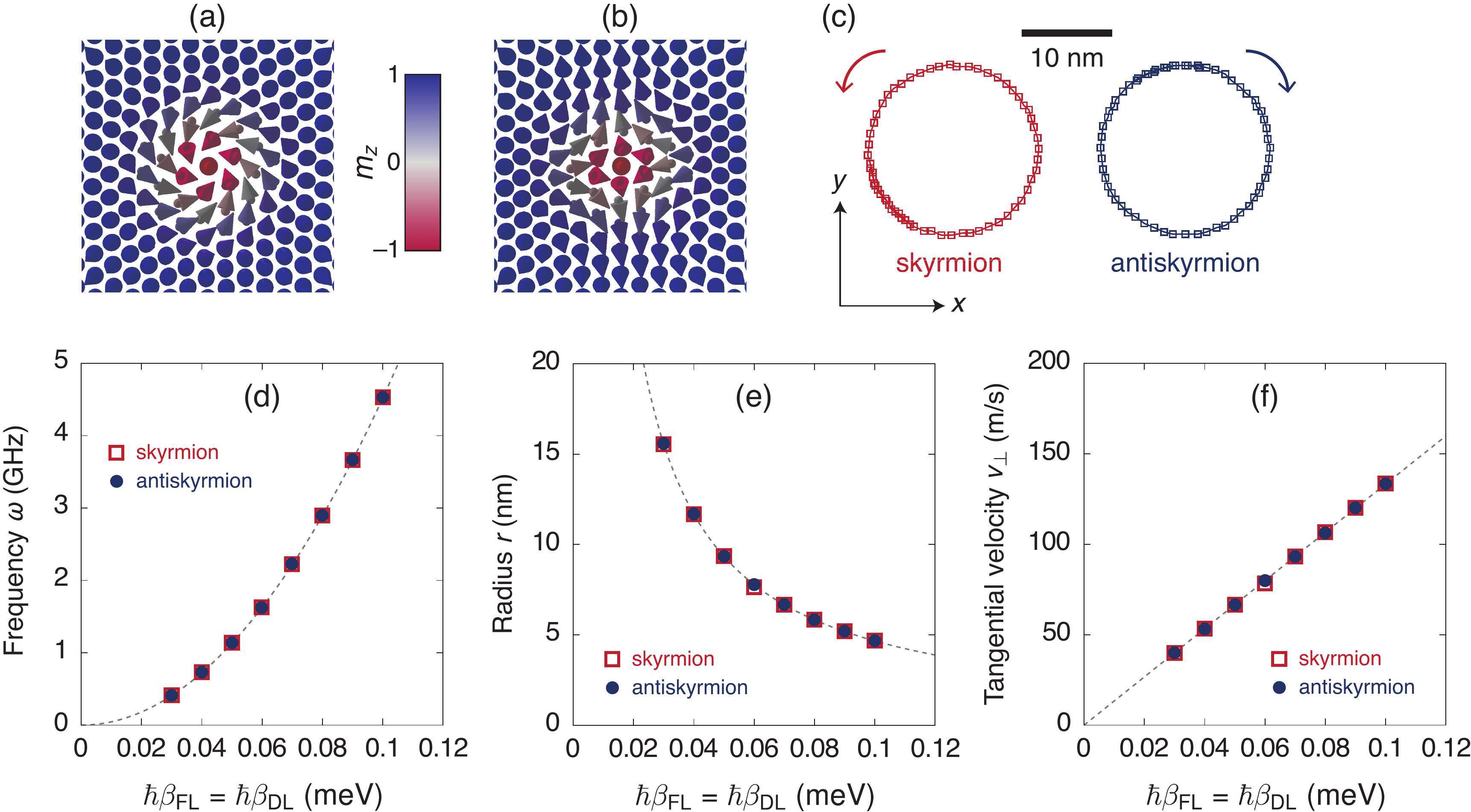}
\caption{Skyrmions and antiskyrmions dynamics without DMI. (a) Equilibrium spin configuration of the skyrmion. (b) Equilibrium spin configuration of the antiskyrmion. (c) Example of trajectories of skyrmion and antiskyrmion motion for $\hbar\beta_\mathrm{FL}=\hbar\beta_\mathrm{DL}=0.05$ meV. (d) SOT dependence of the frequency of the circular motion, where the dashed line represents a quadratic fit. (e) SOT dependence of the radius of the circular motion. (f) Tangential velocity of skyrmions and antiskyrmions as a function of SOT with $\beta_\mathrm{FL}=\beta_\mathrm{DL}$, where the dashed line represents a linear fit.}
\label{fig:zeroDMI}
\end{figure}

The importance of the DMI symmetry can be highlighted further by examining the current-driven dynamics in the absence of DMI altogether. Skyrmions and antiskyrmions remain metastable states because of the frustrated exchange interactions, resulting in the equilibrium profiles shown in Fig.~\ref{fig:zeroDMI}(a) and \ref{fig:zeroDMI}(b).
Since the absence of a chiral interaction results in Bloch and N{\'e}el states being degenerate. $U(\psi) = U_0$ is a constant in Eq.~(\ref{eq:Thielepsi}), so the internal mode $\psi(t)$ becomes a Goldstone mode of the system that can be excited with vanishingly small torques. The Bloch-like skyrmion profile in Fig.~\ref{fig:zeroDMI}(a) is therefore only one possible realization of the metastable state. For a finite deformation $\eta \neq 0$, $\psi(t) = \omega t$ according to Eq.~(\ref{eq:Thielepsi}) and generates a harmonic SOT force in Eq.~(\ref{eq:Thiele}), which results in circular motion. This is confirmed in the spin dynamics simulations as shown in Fig.~\ref{fig:zeroDMI}(c), where circular motion is indeed found with an opposite sense of rotation for opposite topological charges, as expected from the sign of the gyrovector $\mathbf{G}$ in Eq.~\ref{eq:Thiele}. Since the deformation is linear in SOT for the range of values considered, we expect a quadratic variation in the gyration frequency as a function of SOT from Eq.~(\ref{eq:Thielepsi}). This is confirmed in Fig.~\ref{fig:zeroDMI}(d), where the simulated frequencies are well described by a quadratic function. A similar analysis predicts that the radius of gyration should be inversely proportional to the SOT, which is again confirmed by simulations as shown in Fig.~\ref{fig:zeroDMI}(e). Finally, the SOT dependence of the tangential velocity is presented where Fig.~\ref{fig:zeroDMI}(f), where a linear variation is found in agreement with theory. Besides the opposite sense of gyration, these results show that the skyrmion and antiskyrmion trajectories agree quantitatively within the numerical accuracy of the simulations.

These results highlight the rich dynamical behaviour that is possible under SOTs in ultrathin ferromagnetic films, especially for metastable chiral states that are not necessarily the most energetically favourable. The work also links skyrmion dynamics to other known phenomena in micromagnetism, namely Walker breakdown in domain wall motion (trochoidal motion) and vortex core reversal (pair generation). Given the primacy of the DMI symmetry in governing the particle dynamics, our work may spur new avenues of research in materials science where specific surface or interface orientations could be chosen to tailor particular dynamical properties, such as deflected or trochoidal motion, which is absent in most approaches where the focus is on quantifying and controlling rectilinear motion for skyrmion memory and logic applications. The prospect of generating different dynamics with a variety of metastable states within the same material system could also offer new possibilities for studying particle interactions and developing new application paradigms, notably skyrmion generation with a single antiskyrmion `seed'. Recent theoretical work shows that such seeds are likely to appear at finite temperatures~\cite{Boettcher2017} and therefore offer a reliable and efficient means of producing skyrmions and antiskyrmions readily.

\newpage
\subsection*{Methods}

\paragraph{Hamiltonian}
The magnetic Hamiltonian studied is given by
\begin{equation}
	H = -\sum_{\langle ij \rangle} J_{ij} \mathbf{m}_i \cdot \mathbf{m}_j - \sum_{\langle ij \rangle} \mathbf{D}_{ij} \cdot \left(  \mathbf{m}_i \times \mathbf{m}_j \right) - \sum_{i} K \left( \mathbf{m}_i \cdot \hat{\mathbf{z}}\right)^2 - \sum_{i} \mathbf{B} \cdot \mu_\mathrm{s}\mathbf{m}_i,
\label{eq:Hamiltonian}	
\end{equation}
where the first term represents the Heisenberg exchange interaction, the second term the Dzyaloshinskii-Moriya interaction (DMI), the third the uniaxial anistropy along the $z$-axis, and the last term the Zeeman energy associated with an external field $\mathbf{B}$. The indices $\langle ij \rangle$ in the summation for the exchange and DMI terms indicate that single site terms are neglected. The moments are assumed to reside on a hexagonal lattice and $\| \mathbf{m}_i \| = 1$ everywhere. The parameters are extracted from density functional theory (DFT) calculations of the bilayer PdFe system on Ir(111)~\cite{Dupe2014, Dupe2016a}, in which we consider a fcc stacking for the Pd layer. For the Heisenberg exchange,  $J_{ij}$ represents the exchange constant between the magnetic moments $\mathbf{m}_i$ and $\mathbf{m}_j$, where up to 10 nearest-neighbours are taken into account: $J_1 = 14.73$ meV, $J_2=-1.95$ meV, $J_3=-2.88$ meV, $J_4=0.32$ meV, $J_5=0.69$ meV, $J_6=0.01$ meV, $J_7=0.01$ meV, $J_8=0.13$ meV, $J_9=-0.14$ meV, and $J_{10}=-0.28$ meV. We treat the DMI in the nearest-neighbour approximation as shown in Fig.~\ref{fig:geometry}(b) and \ref{fig:geometry}(e), where a magnitude of $1.0$ meV for $\mathbf{D}_{ij}$ is obtained from DFT calculations. The anisotropy constant is $K = 0.7$ meV and we used an applied magnetic field of 20 T along the $z$ direction. The magnetic moment of the Fe atoms is given by $\mu_s = 2.7\mu_\mathrm{B}$, with $\mu_\mathrm{B}$ being the Bohr magneton. For the given parameters the system is in a ferromagnetic ground state close to the transition point to the skyrmion lattice phase, where isolated skyrmions and antiskyrmions can be stabilised. The applied magnetic field is only slightly larger than the critical field $B_c$, with $B=1.06 B_c$.

\paragraph{Atomistic spin dynamics simulations}
The simulation geometry comprises a hexagonal lattice of 100$\times$100 spins with periodic boundary conditions. The ferromagnet is assumed to be one monolayer thick. The dynamics of the spin system described by Eq.~\ref{eq:Hamiltonian} is solved by numerical time integration of the Landau-Lifshitz equation with Gilbert damping and spin-orbit torques given in Eq.~\ref{eq:LLG}. We used a Gilbert damping constant of $\alpha = 0.3$ for all the simulations presented here. The numerical time integration is performed using the Heun method. At the start of each simulation, an equilibrium skyrmion or antiskyrmion profile is first computed by relaxing the system in the absence of the SOT terms. This procedure produces the profiles shown in Figs.~\ref{fig:geometry}(c), \ref{fig:geometry}(d), \ref{fig:geometry}(f), \ref{fig:geometry}(g), \ref{fig:zeroDMI}(a), and \ref{fig:zeroDMI}(b). The simulations are then executed over several ns with a fixed time step in the range of $0.1-10$ fs.

\paragraph{Extension to Thiele model} 
The extension to the Thiele model, expressed by Eq.~\ref{eq:Thielepsi}, is based on the idea that spin-orbit torques (SOT) lead to a significant deformation of the skyrmion/antiskyrmion core. The model is based on two assumptions. First, we assume that all spins in the system are canted toward the film plane under the combined action of the field-like and damping-like SOT. The deformation is assumed to take the form $\mathbf{m} = \mathbf{m}_0 + \eta \, \delta\mathbf{m}$, where the relaxed ground state is $\mathbf{m}_0 = \left(\sin\theta_0 \cos\phi_0, \sin\theta_0 \cos\phi_0, \cos\theta_0\right)$ and
\begin{align*}
\delta m_x &= \cos^2{\theta_0} \cos\phi_0 \cos\left(\phi_0 - \phi_t \right) + \sin\phi_0 \sin\left(\phi_0 - \phi_t \right),  \\
\delta m_y &= \cos^2{\theta_0} \sin\phi_0 \cos\left(\phi_0 - \phi_t \right) - \cos\phi_0 \sin\left(\phi_0 - \phi_t \right), \\
\delta m_z &= \frac{1}{2}\sin{2\theta_0} \cos\left(\phi_0 - \phi_t \right),
\end{align*}
with $\eta$ representing the amplitude of the deformation and $\phi_t$ describing the azimuthal component of the background spins that tilt away from the $z$-axis as a result of the SOT. Second, in addition to the core position $\mathbf{X}(t) = (X(t),Y(t))$, we elevate the helicity parameter $\psi(t)$ to a dynamical variable, which is defined through the azimuthal angle $\phi_0(\mathbf{r},t) = q \tan^{-1}\left[(y-Y(t))/(x-X(t))  \right] + \psi(t)$, where $q=\pm 1$ is the topological charge. Based on this deformation \emph{ansatz}, we derive the equation of motion for $\psi(t)$ using a Lagrangian approach~\cite{Kim2012}, which involves a continuum approximation for the magnetization, $\mathbf{m}(\mathbf{r},t)$, with $\| \mathbf{m} \| =1$. By neglecting coupling terms proportional to $\eta$, we derive the Euler-Lagrange equations leading to Eqs.~(\ref{eq:Thiele}) and (\ref{eq:Thielepsi}), where the  gyrovector term is given by,
\begin{equation}
G_0 = 2\pi \int_{0}^{\infty} dr \; \frac{\partial \theta_0}{\partial r} \sin\theta_0(r),
\end{equation}
the damping factors are
\begin{align}
D_{0} &= \pi \int_{0}^{\infty} dr \; \left[  r \left( \frac{\partial \theta_0}{\partial r}  \right)^2  + \frac{1}{r} \sin^2 \theta_0(r)  \right], \\
D_{\psi} &= 2\pi \int_{0}^{\infty} dr \; r \sin^2 \theta_0(r),
\end{align}
and the SOT efficiency factors are
\begin{align}
\sigma_{0} &= \pi \int_{0}^{\infty} dr \; \left[  r \frac{\partial \theta_0}{\partial r}  + \frac{1}{2} \sin{2 \theta_0(r)}  \right], \\
\sigma_{\psi} &= 2\pi \int_{0}^{\infty} dr \; r \sin^2{\theta_0(r)} \cos{\theta_0(r)}.
\end{align}
Here, the equilibrium (anti-)skyrmion core profile is assumed to possess a cylindrical symmetry, with $r$ being the radial variable in cylindrical coordinates.

Expressions for the helicity-dependent energy, $U(\psi)$, can be found in a similar way by using the continuum approximation of Eq.~(\ref{eq:Hamiltonian}). The dominant contribution comes from the DMI. For the symmetry considered in Fig.~\ref{fig:geometry}(b), we use the form~\cite{Bogdanov2001, Thiaville2012}
\begin{equation}
U_\mathrm{DM} = D \int d^2r \; \left( m_z \frac{\partial m_x}{\partial x} - m_x \frac{\partial m_z}{\partial x} + m_z \frac{\partial m_y}{\partial y} - m_y \frac{\partial m_z}{\partial y}  \right),
\end{equation}
where $D$ is the DMI constant. For skyrmions, we find 
\begin{equation}
U_\mathrm{S}(\psi) = D \, (u_{0,\mathrm{S}} + \eta^2 u_{1, \mathrm{S}}) \cos\psi,
\end{equation}
where
\begin{align}
u_{0,\mathrm{S}} &= 2\pi \int_{0}^{\infty} dr \; \left(r \frac{\partial \theta_0}{\partial r} + \frac{1}{2}\sin2\theta_0(r)  \right), \\
u_{1,\mathrm{S}} &= \pi \int_{0}^{\infty} dr \; \left(r \frac{\partial \theta_0}{\partial r}\cos^2\theta_0(r) + \frac{3}{4}\sin2\theta_0(r) - \frac{1}{8}\sin4\theta_0(r)  \right).
\end{align}
Here, $u_{0,\mathrm{S}}$ is the dominant term, while the deformation-induced contribution $u_{1,\mathrm{S}}$ provides a correction that increases quadratically with the deformation. Only the deformation-induced contribution appears for the antiskyrmion,
\begin{equation}
U_\mathrm{AS}(\psi) = D \, \eta^2 u_{1, \mathrm{AS}} \cos\left(\psi - 2\phi_t\right),
\end{equation}
where
\begin{equation}
u_{1, \mathrm{AS}} = \frac{\pi}{16} \int_{0}^{\infty} dr \; \left( 4 r \frac{\partial \theta_0}{\partial r} \left(1 + 3 \cos{2\theta_0(r)} \right) + 6 \sin{2\theta_0(r)} + \sin{4\theta_0(r)} \right).
\end{equation}
As noted in the main text, an additional energy term $\propto \cos(3\psi)$ is required to describe the atomistic simulations, which arises from discretisation effects due to the underlying hexagonal lattice. This lattice term is not present in the continuum description.

\paragraph{Classification of skyrmion and antiskyrmion trajectories}
The deflected and trochoidal motion for antiskyrmions [with the DMI in Fig.~\ref{fig:geometry}(b)] and skyrmions [with the DMI in Fig.~\ref{fig:geometry}(e)] can involve a wide range of speeds, propagation directions, and gyration frequencies. Classifying these behaviours efficiently from simulation data in order to construct phase diagrams shown in Figs.~\ref{fig:askdyn}(c) and \ref{fig:anisDMI}(c) is therefore a challenging task. We employed algorithms based on machine-learning to classify these trajectories, which were then used with adaptive meshing to identify the different phase boundaries. First, we excluded the annihilation and pair-generation states from the simulation data, which could be identified directly from the magnetization state. Second, the velocity orientations within each simulation run for the remaining data were mapped onto the unit circle, which then served as inputs for classification. The linear motion results in a small cluster of points on the circle, the deflected motion gives a partially filled circle, while the trochoidal motion results in a fully filled circle. A subset of these images (5-10 per state) were then used as learning rules to train the \texttt{Classify} function in the technical computing software \textsc{Mathematica} (version 11.2), which was then used to classify the remaining states. The target resolutions of the phase boundaries in Figs.~\ref{fig:askdyn}(c) and \ref{fig:anisDMI}(c) are 0.01 meV and 0.02 meV, respectively. A brute force search would therefore require 22 500 ($150 \times 150 $) simulation runs for each DMI symmetry, while our iterative method combined with machine learning required only 1831 [Fig.~\ref{fig:askdyn}(c)] and 2736 [Fig.~\ref{fig:anisDMI}(c)] runs, respectively. Given that each simulation run takes 5 to 10 hours of computation time on a single central processing unit (CPU) core, our method provides a more efficient way to explore the parameter space of the dynamical system.

\subsection*{References}
\bibliography{citations}
\bibliographystyle{nature}

\subsection*{Acknowledgements}
This work was partially supported by the Horizon2020 Framework Programme of the European Commission under Grant No. 665095 (MAGicSky). JK acknowledges support from the Deutscher Akademischer Austauschdienst under Award No. 57314019. UR acknowledges support from the Deutsche Forschungsgemeinschaft (grant RI2891/1-1). UR, BD and JS acknowledge the Alexander von Humboldt Foundation, the Deutsche Forschungsemeinschaft (grant DU1489/2-1), the Graduate School Materials Mainz, the ERC Synergy Grant SC2 (No. 610115), the Transregional Collaborative Research Center (SFB/TRR) 173 SPIN+X, and the Grant Agency of the Czech Republic grant no. 14-37427G.

\subsection*{Author Contributions}
BD and SH initiated the project. UR and SvM developed the atomistic spin dynamics code and UR performed the atomistic spin dynamics simulations. UR and JVK interpreted the simulation results and developed the analytical model. SH, BD, JVK and UR wrote the manuscript. All the authors discussed the data.

\subsection*{Competing financial interests}
The authors declare no competing financial interests.

\subsection*{Materials \& Correspondence}
Correspondence and requests for materials should be addressed to Ulrike Ritzmann \\
(\texttt{ulrike.ritzmann@physics.uu.se}).

\newpage
\section*{\large Supplementary Information}
\setcounter{figure}{0}
\renewcommand\thefigure{S\arabic{figure}}
\setcounter{section}{0}
\renewcommand\thesection{S\arabic{section}}
\setcounter{equation}{0}
\renewcommand\theequation{S\arabic{equation}}

\section{Force density}
\label{section:Force_density}

\begin{figure}[b]
\includegraphics[width=0.35\textwidth]{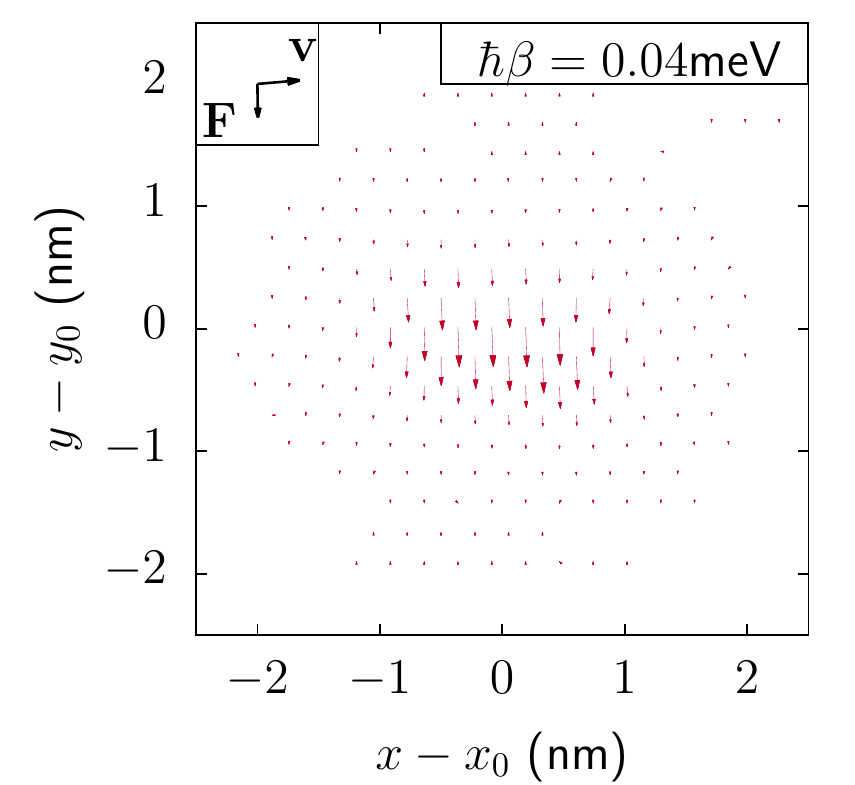}
\hspace{2cm}
\includegraphics[width=0.35\textwidth]{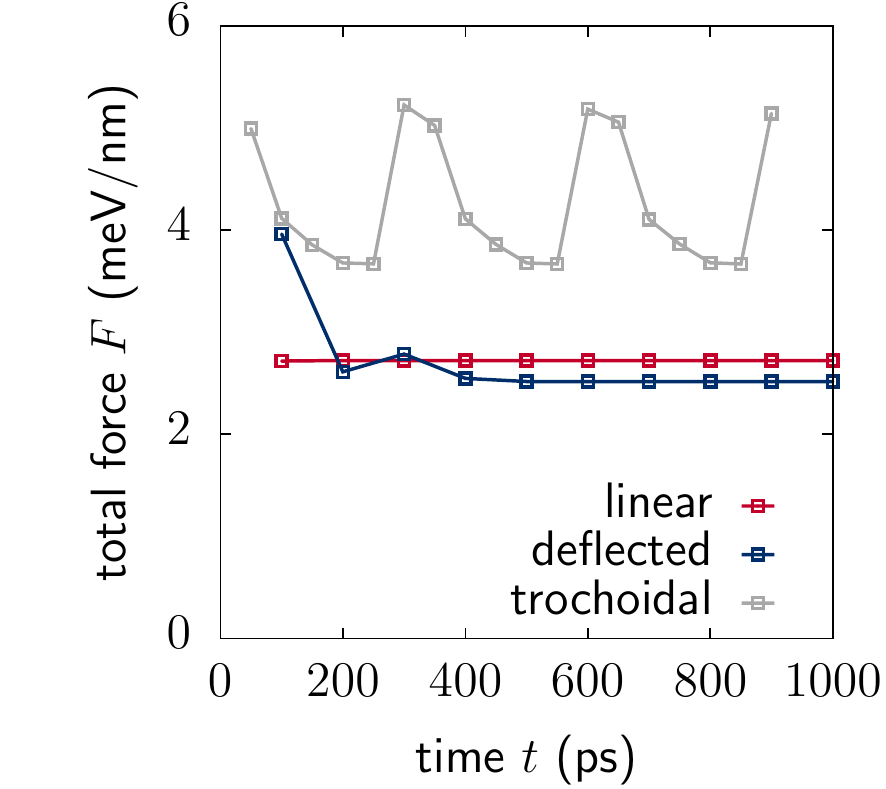}
\caption{Time dependence of the force acting on an antiskyrmion for the different regimes of motion. a) Examplary force density at 200ps in the regime of linear motion for a torque $\hbar\beta_\mathrm{FL}=\hbar\beta_\mathrm{DL} =0.04$meV. The inset shows the direction of the total force. c) Time dependence of the total force for linear motion ($\hbar\beta_\mathrm{FL}=\hbar\beta_\mathrm{DL} =0.04$meV), deflected motion ($\hbar\beta_\mathrm{FL}=\hbar\beta_\mathrm{DL} =0.06$meV) and trochoidal motion ($\hbar\beta_\mathrm{FL}=\hbar\beta_\mathrm{DL} =0.09$meV).}
\label{fig:ftot}
\end{figure}

To obtain further insights in the role of the deformation on the dynamics of skyrmions and antiskyrmions, we calculate the force density $\mathbf{f}(\mathbf{r})$ and the total force $\mathbf{F}=\int_V{\mathbf{f}\mathrm{d}r^2}$ acting on the magnetic texture during the motion following the methodology of Thiele \cite{Thiele1973}. The force density is given by $\mathbf{f}(\mathbf{r})=-\mathbf{H}\cdot \partial\mathbf{m}/\partial\mathbf{r}$, where the field term $\mathbf{H}$ includes contributions from the effective field $\mathbf{B}_\mathrm{eff}$, the Gilbert damping and SOT:
\begin{align}
\mathbf{H}=\mathbf{B}_\mathrm{eff}-\frac{\alpha}{\hbar}\frac{\partial\mathbf{m}}{\partial t}-\beta_\mathrm{FL}\mathbf{P}-\beta_\mathrm{DL}\mathbf{m}\times\mathbf{P} .
\end{align}
In the case of a relaxed state in the absence of SOT terms, the force density is negligible. Internal force contributions from DMI, exchange or the magnetic field are large inside the texture, but all contributions are compensating each other. In the presence of SOT, the force density inside the magnetic texture is no longer vanishing and the motion due to the total force can be described by Eq. (2) in the linear regime. An example of the force density is shown in Fig. \ref{fig:ftot}a) for the linear motion of an antiskyrmion due to SOT with the strength $\hbar\beta_\mathrm{FL}=\hbar\beta_\mathrm{DL}=0.04$meV at a time of $t=2$ps. In that case, the total force points towards $-y$ - direction while the antiskyrmion moves in $x$-direction.

The absolute value of the total force $\mathbf{F}$ dependent on the time is shown in Fig. \ref{fig:ftot}b). In the case of linear motion, the force density remains constant, which results in a constant velocity. In agreement with our analytical model, the force density becomes time-dependent above the threshold for deflected motion. In the regime of the deflected motion, in which the antiskyrmion deforms until it reaches a steady state, the total force decreases at the beginning and remains constant afterwards. For even higher torques in the regime of trochoidal motion, the force is changing periodically in time as illustrated in Fig. \ref{fig:ftot}b) for SOT of the strength of $\hbar\beta_\mathrm{FL}= \hbar\beta_\mathrm{DL}=0.09$meV. In this case, the total force and the speed during the trochoidal motion are linked to each other and change periodically with the same frequency.

\section{Total forces and torques for pair creation}
\begin{figure}
\includegraphics[width=0.45\textwidth]{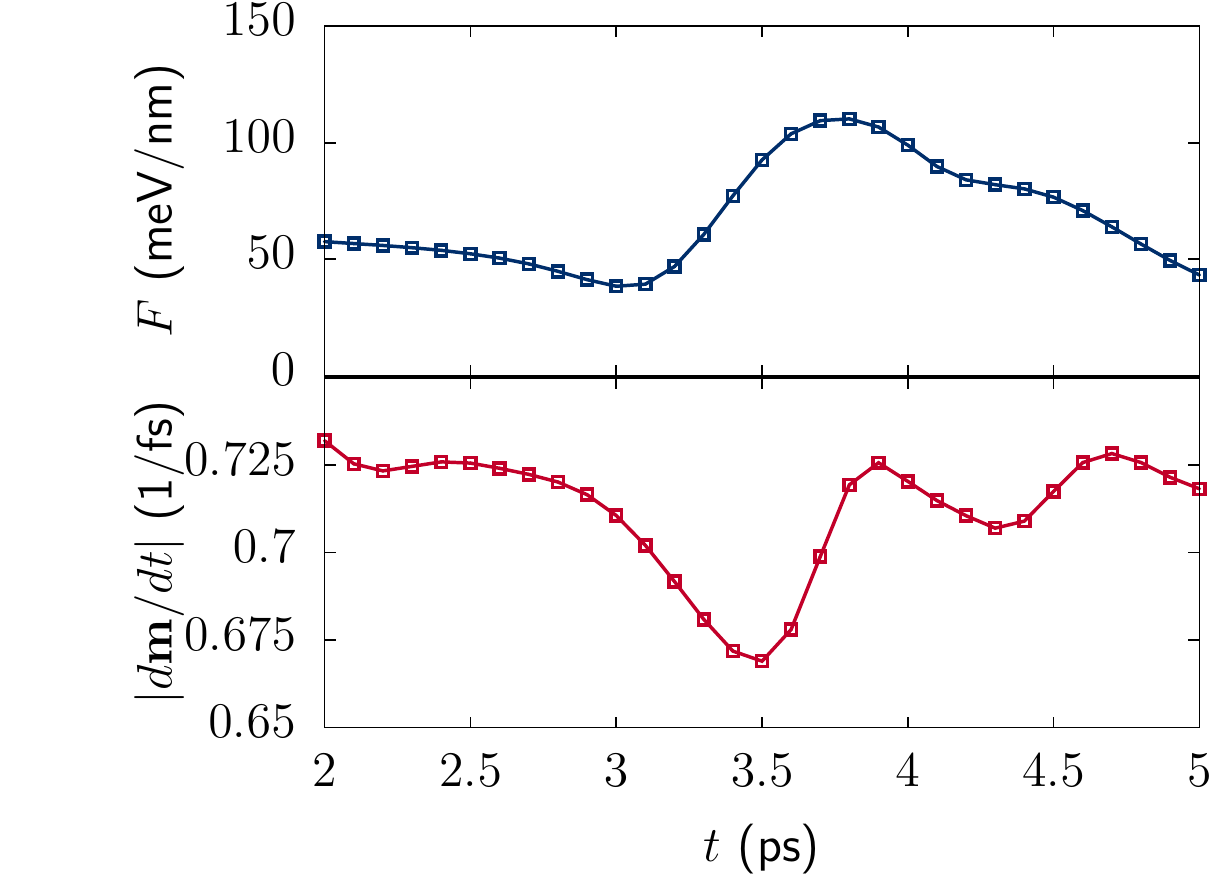}
\caption{Total force and torque during pair creation. The absolute value of the force $|\mathbf{F}|$ (top) and the absolute value of the sum of the acting torques $|\sum_i\partial\mathbf{m_i}/\partial t|$ dependent on the time $t$ are shown exemplary for pair creation at SOT with $\hbar\beta_\mathrm{FL}=0.01$meV and $\hbar\beta_\mathrm{DL}=1.35$meV.}
\label{fig:ftot_pair}
\end{figure}

The rigid body approximation, which is used to describe the motion in the regime of linear, deflected and trochoidal motion, is no longer valid for pair creation. Nevertheless, the calculation of the acting forces can be used to explain the details of the process of pair creation. Fig. \ref{fig:ftot_pair} shows the absolute value of the total force and the sum of the appearing torques $|\partial\mathbf{m}/\partial t|$ for pair creation for SOT with $\hbar\beta_\mathrm{FL}=0.01$meV and $\hbar\beta_\mathrm{DL}=1.35$meV, complementing to the magnetization profiles shown in Fig.~4. A strong deformation of the antiskyrmion is observed in the simulations around 3ps leading to the creation of the pair around 4ps. This is reflected in a clear increase of the total force above 3ps, which then decreases after the pair is formed around 4ps.  
Moreover, we study the sum of the torques acting on the magnetic moments. In the presence of SOT after an initial relaxation of the ferromagnetic background, mainly torque terms are contributing that are acting on the magnetic moment around the texture. In the lower part of Fig. \ref{fig:ftot_pair}, the absolute value of the total torque is shown for pair creation. Here, the torque is reduced during the pair creation process.

\section{Exchange and DMI energy density distribution during the pair creation}

\begin{figure}
\includegraphics[width=0.85\textwidth]{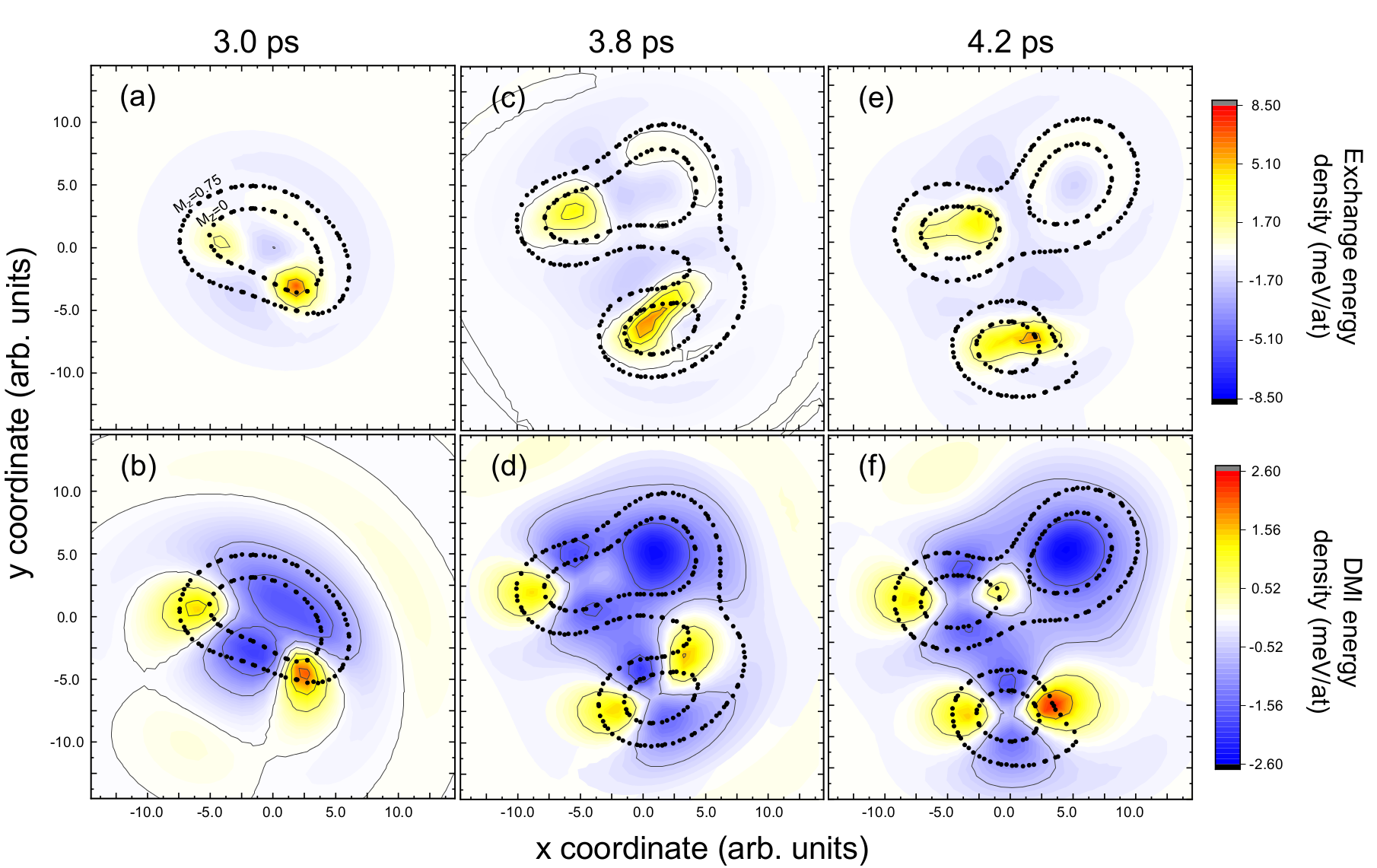}
\label{fig:E_creation}
\caption{Exchange and DMI energy density distribution at 3.0 ps, 3.8 ps and 4.2 ps. The dotted black lines shows lines for constant magnetization $M_z=0.75$ and $M_z=0.0$. The positive DMI energy density (from white to red colors) corresponds to a left rotating non-collinear magnetic texture while the negative energy density (from blue to white) corresponds to a right rotating magnetic texture. Here the SOT favors magnetization aligned with the x direction.}
\end{figure}

To understand the creation of a skyrmion-antiskyrmion from a single antiskyrmion, we show on figure~\ref{fig:E_creation} the exchange and the DMI energy density distribution for three different time: 3.0 ps, 3.8 ps and 4.2 ps corresponding to Fig.~4. The panel (a) shows the exchange energy at 3.0 ps. The skyrmion profile is already non-spherical and the exchange energy shows two distinct maximum. This Exchange energy profile does not correspond to the one of an antiskyrmion which only shows a single maximum in the exchange energy density. The presence of 2 maxima can be viewed as the beginning of the creation process. The panel (b) corresponds to the DMI energy density at 3.0 ps. The DMI energy density corresponds to the profile of an antiskyrmions. The antiskyrmion is composed of left- and right-rotating magnetic regions depending on the axis corresponding to a positive and a negative DMI energy density, respectively. However, the region of left-rotating magnetic texture is showing a lobe and extends away from the antiskyrmion core. This shows that the SOT is enhancing the stability of the left rotating region when is aligns to the polarization of the current.

At 3.8 ps (panel c), the exchange energy shows two distinct maxima which are enclosed within 2 distinct isolines corresponding to $M_z=0.0$. At this time, the DMI energy density distribution exhibits a region characteristic for the antiskyrmion (centered at coordinate (0;-6)) and a region corresponding to a skyrmion-antiskyrmion pair centered at (-2.5;4). The center of the skyrmion corresponds to the negative DMI energy distribution present at (1,5) visible on panel d. The antiskyrmion starts to form below at (-6;2.5) and a right-rotating magnetic texture start to form between this antiskyrmions and the skyrmions.

At 4.2 ps (panel e), the exchange energy density shows 2 regions of positive contribution corresponding to the presence of 2 antiskyrmions and one region of negative contribution corresponding to the skyrmion. The isolines $M_z=0$ are now forming 3 distinct regions in space. The DMI energy density distribution (panel f) confirms the presence of 2 antiskyrmions (exhibiting a butterfly contrast) and one homogeneous negative energy density corresponding to the skyrmion. It is interesting to notice that the left-rotating magnetic regions of the antiskyrmions are aligned but the skyrmion faces one of the right-rotating magnetic region of the antiskyrmion.

\section{Total force density distribution during pair creation}

The previous section describes the evolution of the energy density distribution during the creation of a skyrmion-antiskyrmion pair. In this section, the pair creation is described in term of forces as defined in section~\ref{section:Force_density}.

At 3.0 ps (panel a), the forces are maximum within the antiskyrmion region corresponding to the isolines $M_z=0$. Although the total energy density distribution shows 2 distinct maxima (panel a), the forces show a strong maximum for one half of the deformed antiskyrmion (panel b).

At 3.8 ps, the total energy distribution (panel c) shows two distinct maxima in 2 separate regions of space. The direction of the forces are pointing towards the regions showing a maximum of total energy distribution. This shows that two distinct particle-like magnetic textures are present. The amplitude of the forces (panel d) also shows 2 distinct regions. One is corresponding to the antiskyrmion and shows a strong maximum of the amplitude of the forces density distribution. In the region where the skyrmion-antiskyrmion pair is created, the amplitude of the force starts to rise.

At 4.2 ps, the total energy density shows 3 separated maxima which are enclosed within separate region of space corresponding to the isolines $M_z=0$. These 3 distinct maxima show that 3 particle-like non-collinear magnetic textures are now present. The direction of the forces is interesting because it can characterize the displacement. In this region of the phase diagram, the antiskyrmion have a trochroidal trajectory which means that they should have a non-zero angular momentum. Indeed, the forces are pointing toward the cores of the antiskyrmions $\textbf{r}_0$ which means that $\left( \textbf{r-r}_0 \right) \times \textbf{f} \neq 0$. On the contrary, the skyrmions which have a linear motion, show a force density distribution which is collinear and pointing in the same direction. A closer look on the amplitude of the force density distribution (panel e) shows that there are 3 different regions indicating that the different skyrmions start to get into motion.

\begin{figure}
\includegraphics[width=0.85\textwidth]{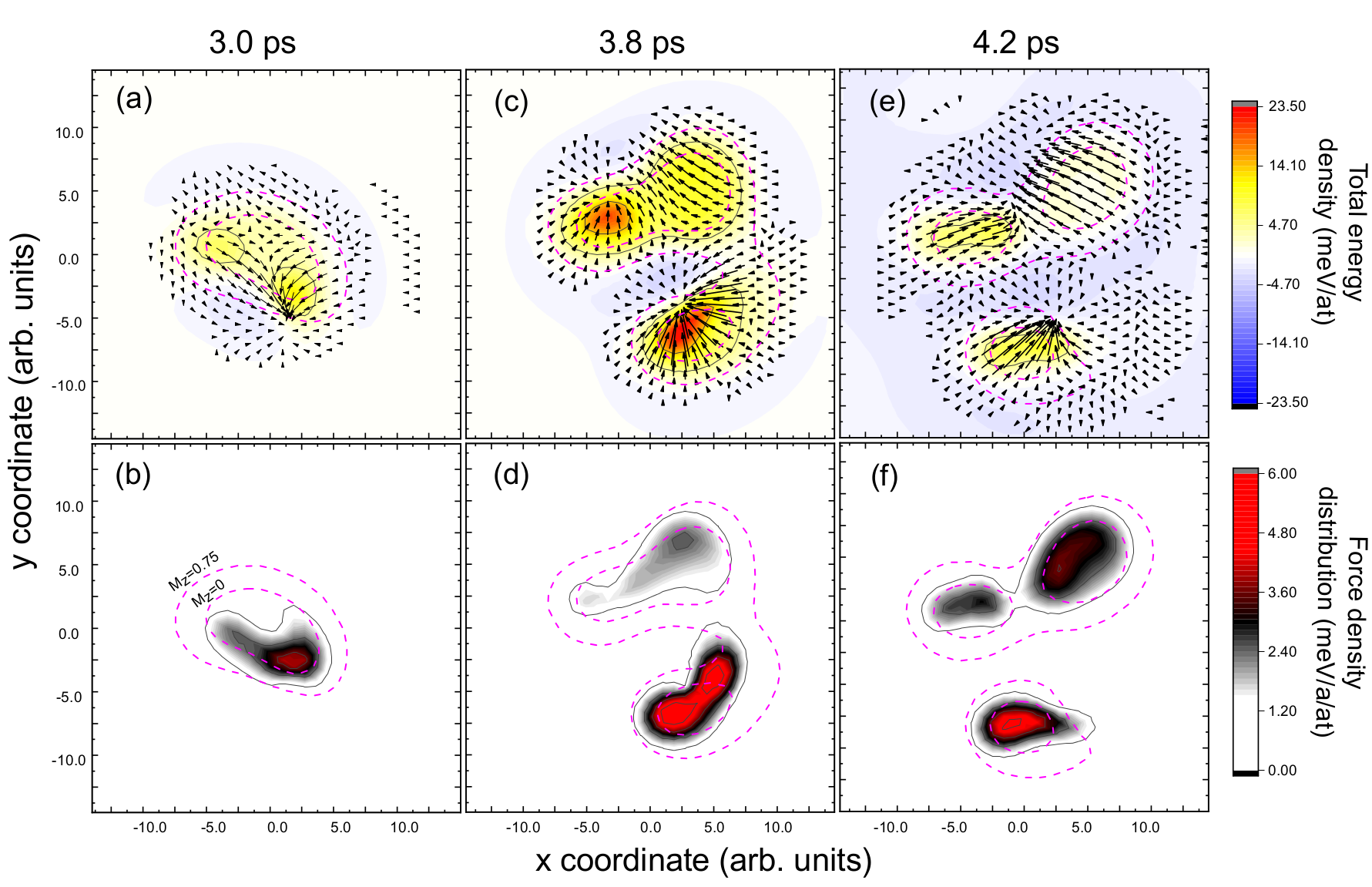}
\label{fig:F_creation}
\caption{Total energy (Exchange, DMI, anisotropy, Zeeman contributions) density distribution and total forces (field- and damping-like) for 3 different simulation times (3.0 ps, 3.8 ps and 4.2 ps) during the pair creation. The magenta lines correspond to the isolines of constant magnetization ($M_z=0$ and $M_z=0.75$). The black arrows of the top panels correspond to the norm (length) and direction of the forces. For better visibility, the norm of the forces is also shown as a color contrast in the bottom panels.}
\end{figure}

\section{Analysis of $U(\psi)$ for antiskyrmions}
\begin{figure}
\centering\includegraphics[width=12cm]{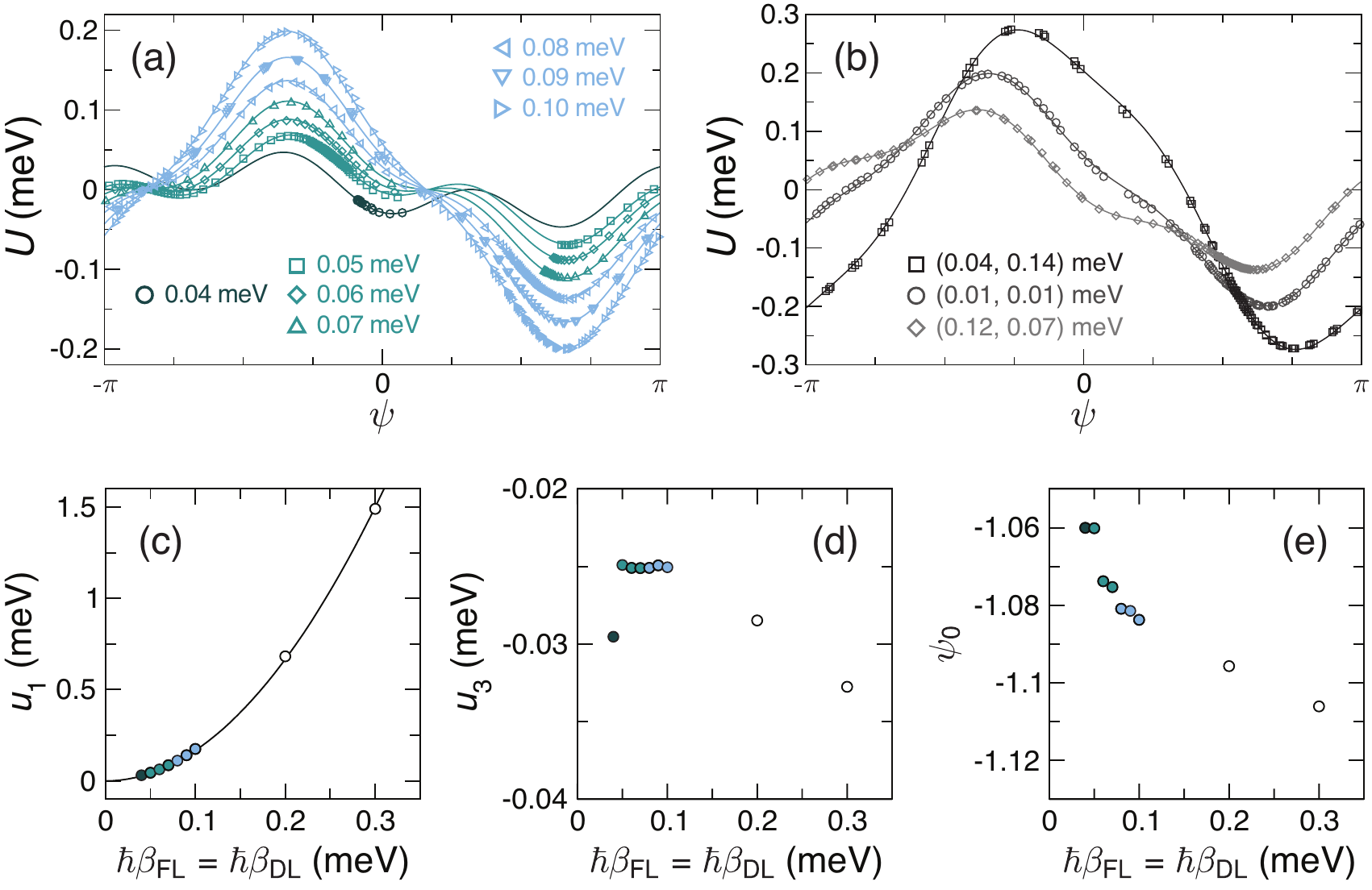}
\caption{Helicity dependence of the antiskyrmion energy, $U(\psi)$. (a) $U(\psi)$ for different values of the SOT with $\beta_\mathrm{FL} = \beta_\mathrm{DL}$. The colour codes refer to the three dynamical regimes: linear (0.04 meV), deflected ($0.05-0.07$ meV), and trochoidal motion ($0.08-0.10$ meV). Points represent values extracted from spin dynamics simulations, while solid lines are fits to Eq.~(\ref{eq:upsi}). (b) $U(\psi)$ for different values of the SOT with $\beta_\mathrm{FL} \neq \beta_\mathrm{DL}$. (c) The coefficients $u_1$ related to the fits in (a). The solid line represents the quadratic function $u_1 = 16.67 (\hbar \beta)^2$. (d) The coefficients $u_3$ related to the fits in (a). (e) The coefficients $\psi_0$ related to the fits in (a).}
\label{fig:Upsi}
\end{figure}

In this section, we present some details of the numerical analysis of the energy $U(\psi)$ extracted from spin dynamics simulations. This discussion complements Fig.~3 of the main text. In Figure~\ref{fig:Upsi}, we present supplementary data for $U(\psi)$ for different values of the SOT terms.
In Fig.~\ref{fig:Upsi}(a) where the case $\beta_\mathrm{FL} = \beta_\mathrm{DL}$ is considered, the data have been fitted with the function
\begin{equation}
U = u_1 \cos(\psi - \psi_0) + u_3 \cos3\psi,
\label{eq:upsi}
\end{equation}
where the first term is the deformation-induced term that depends on the DMI and the second term is a lattice term, as discussed in the main text. As the SOT is increased, we observe that the energy landscape becomes predominantly governed by the cosine term, as expected from the theory. We note that the energies are well described by this function for the values of the SOT considered. Moreover, the position of the global extrema do not depend on the SOT, which reflects the fact that the overall tilt $\phi_t$ varies little along the contour $\beta_\mathrm{FL} = \beta_\mathrm{DL}$. A shift in the positions of the extrema, along with the overall shape of the energy landscape, can be seen for cases where $\beta_\mathrm{FL} \neq \beta_\mathrm{DL}$ as shown in Fig.~\ref{fig:Upsi}(b). This is consistent with the extended Thiele model.

The SOT dependence of the coefficients $u_1$, $u_3$, and $\psi_0$ for $\beta_\mathrm{FL} = \beta_\mathrm{DL}$ is shown in Figs.~\ref{fig:Upsi}(c)-(e). $u_1$ is found to be well described by a quadratic function, which is consistent with the model prediction
\begin{equation}
u_1 = D \, u_{1, \mathrm{AS}} \, \eta^2
\end{equation}
if we assume that the deformation $\eta$ is linear in SOT. Here $D$ is the Dzyaloshinskii-Moriya coefficient and $u_{1, \mathrm{AS}}$ is a numerical prefactor that depends on the details of the shape of the antiskyrmion core (see Methods section of the main text). We note that the linear dependence of $\eta$ on the SOT can be expected from simple field arguments and is consistent with the behaviour discussed in Fig.~6 of the main text. The lattice term $u_3$ is found to be largely independent of the SOT, with deviations only seen at small (0.04 meV) and larger values (0.2, 0.3 meV) of the SOT. The discrepancy for the former may arise from uncertainties in the fit, since the linear motion does not lead to a large range of $\psi$ values being sampled. For the larger values, we note that SOT-induced core deformations become more important, where the core itself loses much of its cylindrical symmetry. Nevertheless, the general trends are consistent with the model. Similarly, the fitted coefficient $\psi_0$ is also found to be largely independent of the SOT, which is consistent with the fact that $\phi_t$ varies little with the SOT for $\beta_\mathrm{FL} = \beta_\mathrm{DL}$.

\section{Pair generation and skyrmion crystallisation}
\begin{figure}
\centering\includegraphics[width=14cm]{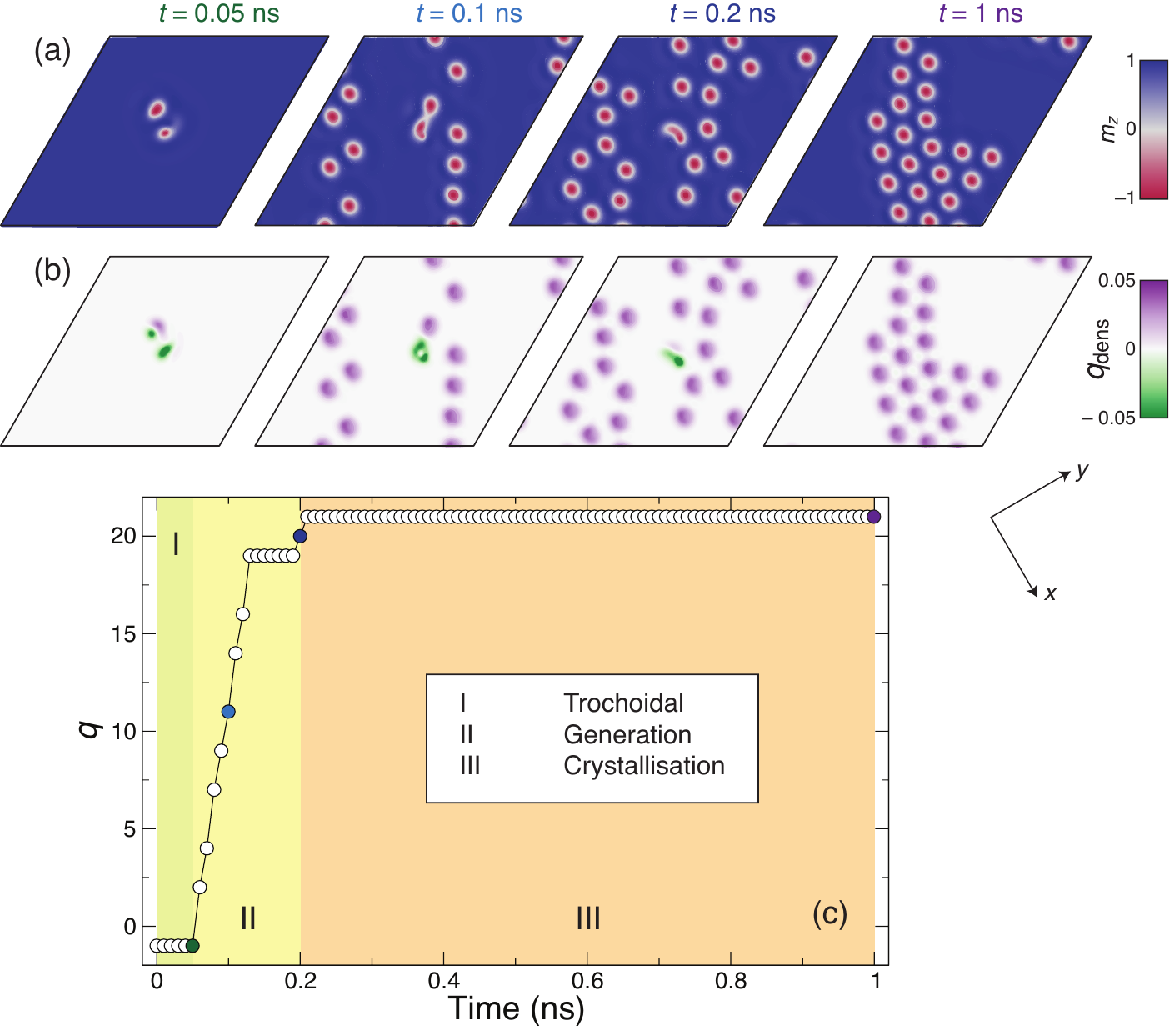}
\caption{Skyrmion-antiskyrmion pair generation and skyrmion crystallisation for $\hbar \beta_\mathrm{FL} = 0.13$ meV and $\hbar \beta_\mathrm{DL} = 1.4$ meV. Snapshots of the (a) $m_z$ component and (b) topological charge density $q_\mathrm{dens}$, which are approximately centred on the position of seed antiskyrmion. (c) Total topological charge $q$ as a function of time, where the snapshots in (a,b) and the three phases, trochoidal motion, pair generation, and skyrmion crystallisation, are indicated.}
\label{fig:skyxtal}
\end{figure}

In this section, we provide further details of the skyrmion-antiskyrmion pair generation, as shown in Fig.~4 in the main text. In Fig.~\ref{fig:skyxtal}, we illustrate a particular example in which the generation leads to a skyrmion `crystallite' that condenses from a dilute gas of skyrmions.
Here, the SOT is dominated by the damping-like term ($\hbar \beta_\mathrm{FL} = 0.13$ meV, $\hbar \beta_\mathrm{DL} = 1.4$ meV). The initial state of the simulation comprises an antiskyrmion at the centre of the simulation grid. Under the applied SOT, the antiskyrmion undergoes a trochoidal motion, leading first to a large core deformation and then a nucleation of the skyrmion-antiskyrmion pair, as shown at $t = 0.05$ ns in Figs.~\ref{fig:skyxtal}(a) and \ref{fig:skyxtal}(b). This pair then separates under the action of the SOT and is followed by the periodic generation of other pairs. Interestingly, the nucleated antiskyrmions do not appear to escape the region of generation and are annihilated more readily than their skyrmion counterparts, leading to an excess in the skyrmion population as shown at $t = 0.1$ and $0.2$ ns. This is also quantified in Fig.~\ref{fig:skyxtal}(c), where the total topological charge $q$ is shown as a function of time. An event occurs after $t=0.2$ ns that annihilates the seed antiskyrmion, resulting in a dilute gas of pure skyrmions. Because of the attractive interaction made possible by the frustrated exchange, this gas progressively condenses into a `crystallite' as time evolves, where successive collisions lead to the ordered configuration shown at $t = 1$ ns in Fig.~\ref{fig:skyxtal}.

\section{SOT-induced tilts and model for core deformation}
\begin{figure}
\centering\includegraphics[width=10cm]{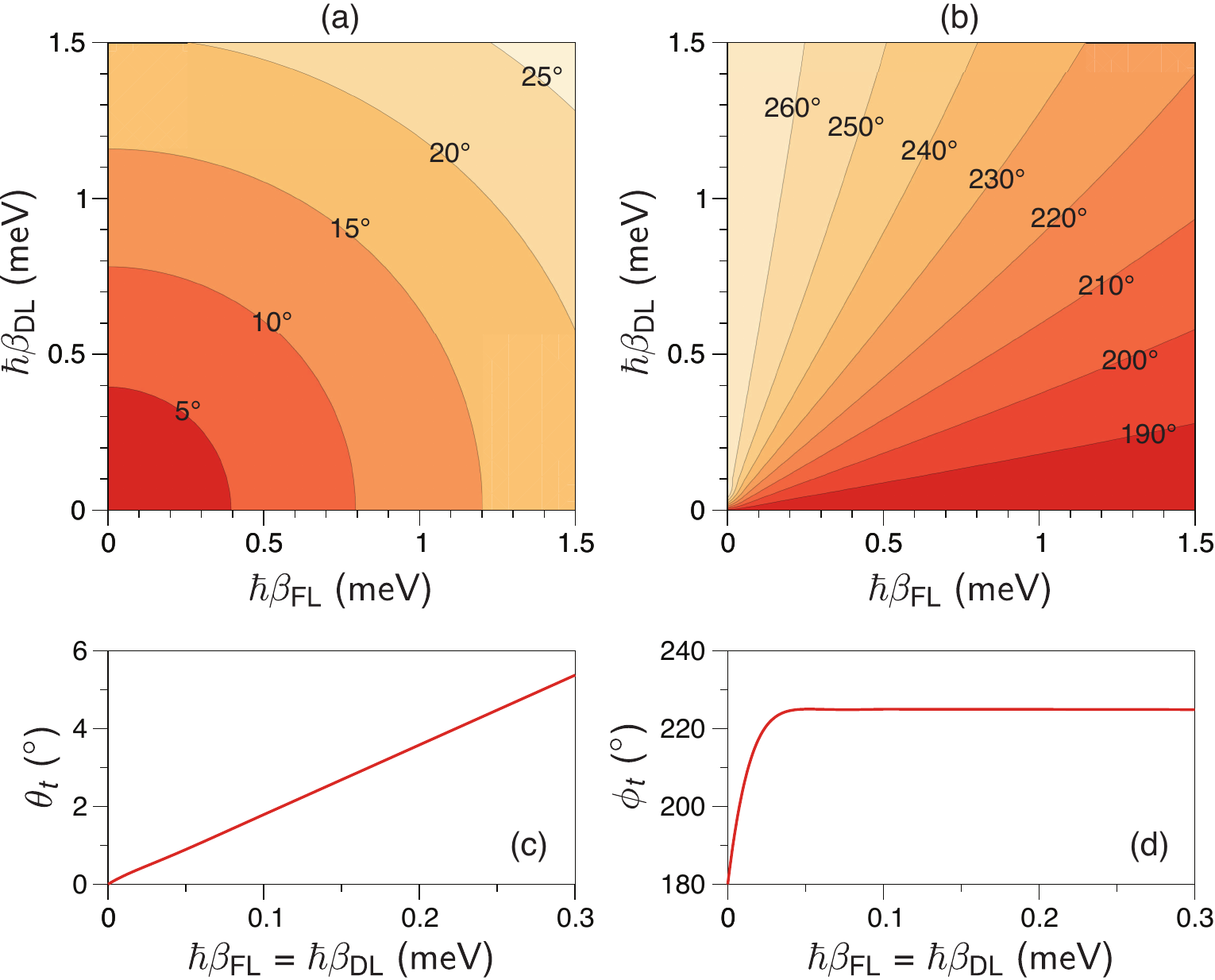}
\caption{SOT-induced tilts on the uniform background magnetization. (a) Contour plot of the polar tilt angle, $\theta_t$ as a function of SOT. (b) Contour plot of the azimuthal tilt angle, $\phi_t$, as a function of SOT. (c) $\theta_t$ as a function of the SOT for $\beta_\mathrm{FL}  = \beta_\mathrm{DL}$. (d) $\phi_t$ as a function of the SOT for $\beta_\mathrm{FL}  = \beta_\mathrm{DL}$.}
\label{fig:tilts}
\end{figure}

In this section, we describe the magnetization tilts induced by the spin-orbit torques and how these are used to motivate the \emph{ansatz} for the core deformation profile. The influence of the field-like and damping-like torques on the uniform ferromagnetic state is presented in Fig.~\ref{fig:tilts}.
The equilibrium configuration was computed for different values of the SOT (with $\mathbf{P} = \hat{\mathbf{x}}$) in the absence of a skyrmion or antiskyrmion with the magnetic Hamiltonian given in the Methods section of the main text. Overall, the combined action of the field-like and damping-like torques can be assimilated to the presence of a fictitious SOT magnetic field applied in the film plane, which tilts the magnetization away from the $z$-axis, characterised by the polar angle $\theta_t$, along an azimuthal orientation given by the angle $\phi_t$ taken from the $x$-axis in the usual way. We note that the polar tilt is largely independent of the $\beta_\mathrm{DL}/\beta_\mathrm{FL}$ ratio, which is expected for an applied magnetic field in the film plane [Fig.~\ref{fig:tilts}(a)]. On the other hand, the azimuthal tilt depends primarily on this ratio, rather than the magnitudes of the SOT terms, which can also be expected from this field argument [Fig.~\ref{fig:tilts}(b)]. The polar tilt is found to vary linearly as a function of SOT for the range of values considered [Fig.~\ref{fig:tilts}(c)], while the azimuthal tilt is largely independent of the SOT strength above a threshold for the $\beta_\mathrm{FL}  = \beta_\mathrm{DL}$ case [Fig.~\ref{fig:tilts}(d)].

Our \emph{ansatz} for the core deformation is based on this idea that the SOT acts like an effective magnetic field on the equilibrium state. We first discuss the macrospin approximation. Let $\mathbf{m}$ be a unit vector representing the magnetization orientation, which we can write in terms of the spherical polar angles in the usual way, $\mathbf{m} = \left( \sin\theta \cos\phi, \sin\theta \sin\phi, \cos\theta \right)$. Let us consider the effect of an applied magnetic field in the film plane, $h = h_0 \left( \cos\phi_h, \sin\phi_h, 0 \right)$. Under the action of the Landau-Lifshitz damping torque, the time rate of change in $\mathbf{m}$ can be written as
\begin{equation}
\frac{d \mathbf{m}}{dt} = -\alpha \mathbf{m} \times \left(\mathbf{m} \times \mathbf{h} \right).
\end{equation}
By assuming $\theta=\theta(t)$ and $\phi=\phi(t)$, this leads to the two linearly independent equations,
\begin{align}
\frac{d\theta}{dt} &= \alpha h_0 \cos\theta \cos\left(\phi-\phi_h \right), \\
\frac{d \phi}{dt} &= -\alpha h_0 \frac{\sin\left(\phi-\phi_h \right)}{\sin\theta}.
\end{align}
Therefore, we can assume that applied in-plane fields will lead to tilts of the form
\begin{align}
\delta\theta &=  \eta \cos\theta \cos\left(\phi-\phi_h \right), \label{eq:thetatilt} \\
\delta\phi &= -\eta \frac{\sin\left(\phi-\phi_h \right)}{\sin\theta}, \label{eq:phitilt}
\end{align}
where $\eta$ is a parameter that describes the tilt amplitude. We then extend this approach to describe the tilting of any arbitrary magnetization field $\mathbf{m} = \mathbf{m}(\mathbf{r},t)$ by assuming that $\delta\theta = \delta\theta(\mathbf{r},t)$ and $\delta\phi = \delta\phi(\mathbf{r},t)$. Let us consider deformations about an equilibrium configuration, $(\theta_0,\phi_0)$, due to the same uniform in-plane applied field. To linear order in the deformations, the magnetization field can be written as
\begin{align*}
m_x(\mathbf{r},t) &= \sin\theta_0 \cos\phi_0 + \delta\theta \cos\theta_0 \cos\phi_0 - \delta\phi \sin\theta_0 \sin\phi_0, \\
m_y(\mathbf{r},t) &= \sin\theta_0 \sin\phi_0 + \delta\theta \cos\theta_0 \sin\phi_0 + \delta\phi \sin\theta_0 \cos\phi_0, \\
m_z(\mathbf{r},t) &= \cos\theta_0 - \delta\theta \sin\theta_0.
\end{align*}
By using the expression for the tilts in Eqs.~(\ref{eq:thetatilt}) and (\ref{eq:phitilt}), we obtain
\begin{align}
m_x(\mathbf{r},t) &= \sin\theta_0 \cos\phi_0 + \eta \left(\cos^2\theta_0 \cos\left(\phi-\phi_h \right) \cos\phi_0  +    \sin\left(\phi-\phi_h \right) \sin\phi_0  \right), \\
m_y(\mathbf{r},t) &= \sin\theta_0 \sin\phi_0 + \eta \left(\cos^2\theta_0 \cos\left(\phi-\phi_h \right) \sin\phi_0 -    \sin\left(\phi-\phi_h \right) \cos\phi_0  \right), \\
m_z(\mathbf{r},t) &= \cos\theta_0 - \frac{1}{2} \eta \sin\left(2\theta_0\right)\cos\left(\phi-\phi_h \right),
\end{align}
which is the model described in the main text, where we have substituted the SOT polarization angle $\phi_p$ in the place of $\phi_h$ here. Note that this deformation \emph{ansatz} does not conserve the length of the magnetization vector,
\begin{equation}
\| \mathbf{m} (\mathbf{r},t) \|^2 \simeq 1 + \frac{3}{4}\eta^2.
\end{equation} 
This means that $\mathbf{m}$ needs to be rescaled when making comparisons with simulation data.

\begin{figure}
\centering\includegraphics[width=16cm]{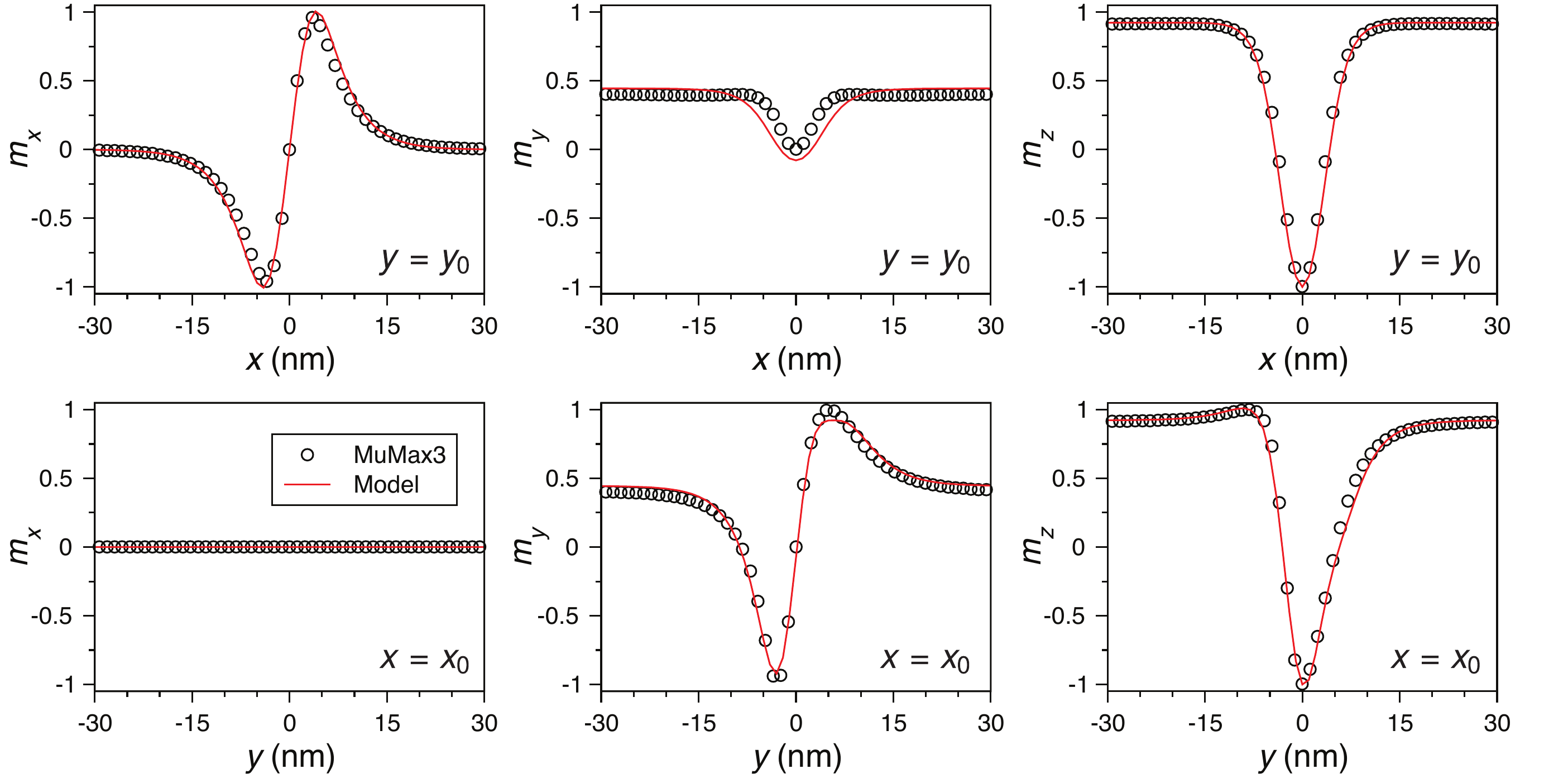}
\caption{Comparison of the deformation \emph{ansatz} (lines) with the magnetization profile of a skyrmion computed with micromagnetics simulations (circles) for an in-plane field of $\mu_0 H_y = 400$ mT.}
\label{fig:deformprofile}
\end{figure}

\begin{figure}
\centering\includegraphics[width=12cm]{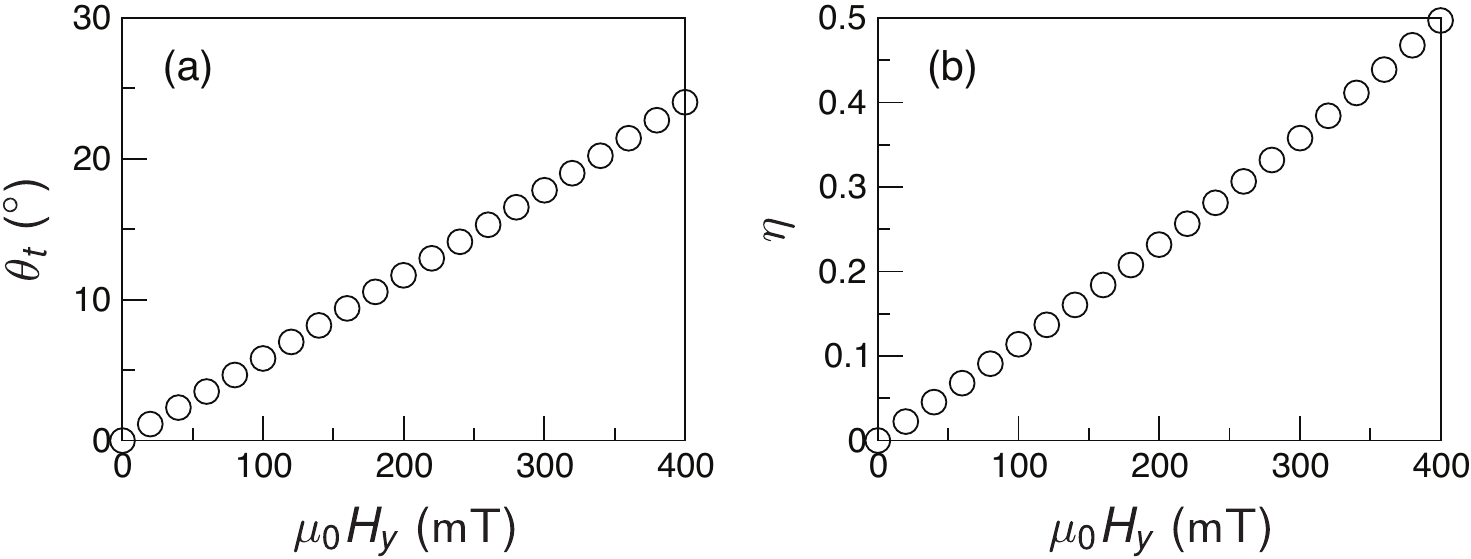}
\caption{(a) Background magnetization tilt, $\theta_t$, and (b) deformation amplitude, $\eta$, as a function of applied magnetic field along the $y$ axis.}
\label{fig:coredata}
\end{figure}

To explore the validity of this \emph{ansatz}, we performed micromagnetics simulations of a static skyrmion in the presence of an applied field along the $y$ direction. Micromagnetics simulations were used (rather than the atomistic code) as they provide a better correspondence with the continuum approximation in which this deformation model is developed. We used the \texttt{MuMax3} code~\cite{Vansteenkiste2014} and considered a $300 \times 300 \times 0.6$ nm film that was discretised with $256 \times 256 \times 1$ finite difference cells. Periodic boundary conditions were used. We assumed an exchange constant of $A =  16$ pJ/m, a uniaxial anisotropy of $K = 1.3$ MJ/m$^3$, a saturation magnetization of $M_s = 1.1$ MA/m, and an interfacial DMI of $D = 2.5$ mJ/m$^2$. The relaxed skyrmion profile can be well described with the double soliton model~\cite{Braun1994, Romming2015},
\begin{equation}
m_z(r) = 1- \frac{4 \cosh^2{c}}{\cosh{2c}+ \cosh{\left(2 r/\Delta \right)}},
\label{eq:doublesoliton}
\end{equation}
where $r$ is the radial coordinate, $c$ is a measure of the skyrmion radius, and $\Delta = \sqrt{A/K_0}$ is the wall width parameter. This profile was used to define the functional form of the polar angle $\theta_0(r)$ in the model. An example of fits to the simulation data with the deformation \emph{ansatz} is presented in Fig.~\ref{fig:deformprofile}.
We observe excellent quantitative agreement between the deformation described by the model and the micromagnetics simulations for all three components of the magnetization. From the fits, we can correlate the deformation amplitude, $\eta$, with the background magnetization tilt, $\theta_t$. This is shown in Fig.~\ref{fig:coredata}, where these parameters are shown as a function of applied in-plane magnetic field.
From this data, we deduce the empirical relationship, $\eta \simeq 0.021 \theta_t$, where the tilt angle $\theta_t$ is expressed in degrees. These results also suggest that we can expect the deformation $\eta$ to be a linear function of the SOT strength, since the SOT plays a similar role to a magnetic field for a static profile.

\end{document}